\documentclass[preprint,12pt]{elsarticle}

\usepackage{amssymb}

\journal{Nuclear Physics A}

\begin{document}

\begin{frontmatter}



\title{Structure of the ground and excited states in $_{\Lambda}^{9}$Be nucleus}

\author{A. V. Nesterov}
\ead{nesterov@bitp.kiev.ua}

\author{Yu. A. Lashko\corref{cor1}}
\ead{ylashko@gmail.com}\cortext[cor1]{}

\author{V. S. Vasilevsky}
\ead{vsvasilevsky@gmail.com}

\address{Bogolyubov Institute for Theoretical Physics,\\
 Kiev 03143, Ukraine
}


\date{\today }

\begin{abstract}
We investigate properties of bound and resonance states in the $_{\Lambda}^{9}$Be nucleus.
To reveal the nature of these states, we use a three-cluster $2\alpha+\Lambda$ microscopic model.
The model incorporates Gaussian and oscillator basis functions and reduces a three-cluster Schr\"{o}dinger equation to a two-body like many-channel problem with the two-cluster subsystems ($_{\Lambda}^{5}$He and $^8$Be) being in a bound or a pseudo-bound state. Influence of the cluster polarization on the energy and widths of resonance states in $_{\Lambda}^{9}$Be and on elastic and inelastic $_{\Lambda}^{5}$He+$\alpha$ scattering is analysed.

\end{abstract}




\end{frontmatter}



\section{Introduction}

We apply a microscopic three-cluster model to study the hypernucleus
$_{\Lambda}^{9}$Be. This nucleus is considered as a three-cluster system
$\alpha+\alpha+\Lambda$. Our aim is to examine both discrete and continuous
spectrum states of $_{\Lambda}^{9}$Be. This research is performed within a
microscopic three-cluster model referred as AMGOB (the Algebraic Model \ of scattering with
the Gaussian and Oscillator Bases). This model was formulated in Ref. \cite{2009NuPhA.824...37V}.
In Refs. \cite{2009NuPhA.824...37V, 2009PAN....72.1450N, 2014UkrJPh..59.1065N, 2017NuPhA.958...78L,
2017UkrJPh..62..461V}, the AMGOB\ has been successfully applied to study
structure of bound and resonance states in the light nuclei $^{7}$Be, $^{7}%
$Li, $^{8}$Li, $^{8}$B, $^{10}$Be and $^{10}$B. The model has been also
applied in Ref. \cite{2012PAN.75.818V} to study the astrophysical $S$ factors
of the capture reactions of astrophysical importance.
By this reason, the model is particularly appealing for investigating
different\ two-body decay channels of the compound hypernucleus.

The energy of $1/2^+$ ground state in $_{\Lambda}^{9}$Be is -3.12 MeV with respect to its lowest binary decay threshold $_{\Lambda}^{9}$Be$\rightarrow^5_\Lambda$He$+\alpha$ and -6.63 MeV relative to its three-cluster $2\alpha+\Lambda$ threshold. Unlike $^9$Be, which is a Borromean nucleus, $_{\Lambda}^{9}$Be has a bound two-body subsystem $_{\Lambda}^{5}$He. That is why it is important to take into account the possibility for
the $_{\Lambda}^{9}$Be hypernucleus to decay via $_{\Lambda}^{9}$Be$\rightarrow^5_\Lambda$He$+\alpha$ channel.
The ground state of $^8$Be subsystem is known to be a very narrow resonance just near $2\alpha$ threshold. Hence space correlations between $\alpha$-particles should also be considered properly.
AMGOB model gives us a possibility to take into account two coupled binary cluster configurations $^5_\Lambda$He$+\alpha$ and $^8$Be$+\Lambda$ allowing for  $^5_\Lambda$He and $^8$Be to be polarized.
The term "cluster polarization" connotes changing energy of a two-cluster subsystem (and, hence, change of its shape and/or size) due to the interaction with the third cluster.

The light hypernuclei have been investigated within different models in Refs.
\cite{2009PrPNP..63..339H,
2003PrPNP..51..223H,
2018FrPhy..13.2106H, 2018PhRvC..97c4324K,
2018PhRvC..97b4330K, 2018PhRvC..97f4315W,
2014EPJWC..6609013M, 2009PhRvC..80e4321H,
1999JPhG...25..961P, 2000PhRvL..85..270H, 1988PhRvC..38..854Y, 2014PTEP.2014k3D01F,
2002PhRvL..88h2501A, 2006PrPNP..57..564H,
2015PhRvC..92d4326I, 2018PhRvC..97c4302L,
2011PhRvC..83d4323I, 2015RPPh...78i6301F,
2016NuPhA.954..260V, 2005NuPhA.753..233T,
2004PhRvC..70b4002F, 2004PhRvC..70d7002F, 1997PThPh..97..881H}.
In Ref. \cite{2018PhRvC..97b4330K} $L^\pi=0^{+}$ ground state and $L^\pi=2^{+}$ excited states
of $_{\Lambda}^{9}$Be have been investigated with a microscopic
cluster model. Spin of a $\Lambda$-hyperon was disregarded in \cite{2018PhRvC..97b4330K} and, hence,
all the states of $_{\Lambda}^{9}$Be hypernucleus have been classified with the values of the total orbital momentum $L$. For the description of the core nucleus $^{8}$Be the generator coordinate method of a microscopic $2\alpha$ cluster model has been applied. The $\Lambda$-nucleus potentials have been constructed
by folding $\Lambda N$ interactions with the nuclear density calculated by the
microscopic cluster model. A core polarization has been taken into account by
artificial enhancing the central part of $NN$-potential. In this assumption,
strengthening of the effective central nuclear interactions acts like an
intensification of the inter-cluster potentials. This procedure simulates an
additional attraction to the nuclear system from $\Lambda N$ interaction. The
optimum value of the enhancement factor was chosen to minimize the energy of
the total system. Authors claimed that particularly remarkable core
polarization effects are found in $_{\Lambda}^{9}$Be, because $^{8}$Be is a
very fragile system of the quasi-bound $2\alpha$ state. The core polarization
effects have been seen in the nuclear size change and the energy changes caused by
the $\Lambda$-particle in $_{\Lambda}^{9}$Be. The significant shrinkage of
$2\alpha$ structure in $_{\Lambda}^{9}$Be has also been reported.

In Ref. \cite{2019FBS....60...30L} energy spectra of bound and resonance states of
$_{\Lambda}^{9}$Be have been calculated within the framework of $\alpha
+\alpha+\Lambda$ three-body model. The $\alpha-\alpha$ interaction was chosen
so to reproduce the observed $\alpha-\alpha$ scattering phase shift and the
ground state of $^{8}$Be within the $\alpha-\alpha$ orthogonality condition
model. The $\Lambda\alpha$ interaction was obtained by folding the $\Lambda N$
interaction into the $\alpha$ cluster wave function. Even- and odd-states of
$\Lambda N$ interaction have been adjusted so as to reproduce the observed
binding energies of the ground states in $_{\Lambda}^{5}$He and $_{\Lambda}^{9}%
$Be. For the resonant states of $_{\Lambda}^{9}$Be the complex scaling method
has been employed. The level structure has been categorized into $^{8}%
$Be-analogue states, genuine hypernuclear states, $^{9} $Be analogue states,
which have already been discussed in \cite{1985PThPS..81...42M,%
 1983PThPh..69..918B, 1985PThPS..81..147I} and some new states
located at more than 10 MeV above the $\alpha+\alpha+\Lambda$ threshold states.

An extensive discussion of the structure of genuine hypernuclear states of
$_{\Lambda}^{9}$Be, as well as $^{8}$Be$^{\ast}$-analogue states, within a
microscopic $\alpha+\alpha(\alpha\ast)+\Lambda$ cluster model is given also in
review paper \cite{2009PrPNP..63..339H}.

In Ref. \cite{2000PAN....63..336F} $_{\Lambda}^{9}$Be hypernucleus has been treated
as the $S=1/2$, $T=0$ bound state of the three-cluster system $\alpha
\alpha\Lambda$. The cluster-reduction method is used to solve the s-wave
differential Faddeev equations. Phenomenological potentials have been used to
describe $\Lambda\alpha$ and $\alpha\alpha$ interactions. The authors have
considered boundary-value problems corresponding to the bound states in the
$\alpha\alpha\Lambda$ system and the problems of low-energy alpha-particle
scattering on a $_{\Lambda}^{5}$He hypernucleus. The s-wave phase shift for
$\alpha-_{\Lambda}^{5}$He scattering has been shown to behave anomalously at energies of relative motion
below 1 MeV being small and positive. The scattering length has been observed to be large in magnitude and negative, which has been attributed by the authors to the presence of a virtual level in the $\alpha
\alpha\Lambda$ system near the threshold for scattering.

In \cite{2014PhRvL.113s2502W} the first ab initio calculations for p-shell
single-$\Lambda$ hypernuclei using no-core shell model approaches with
explicit hyperons have been presented. In addition to chiral
two- and three-nucleon interactions, they used leading-order (LO) chiral
hyperon-nucleon (YN) interactions and a meson-exchange hyperon-nucleon
interaction. They have shown that the chiral hyperon-nucleon interactions
provide the ground-state and excitation energies that generally agree with
experiment within the cutoff dependence. At the same time they demonstrated
that hypernuclear spectroscopy provides tight constraints on the
hyperon-nucleon interactions.
A peculiarity of $_{\Lambda}^{9}$Be is that the spin-doublet resulting from the
$2^{+}$ state in $^{8}$Be is practically degenerate, with the higher $J$ state
being at slightly lower excitation energy experimentally, contrary to
other light hypernuclei. The LO chiral YN interactions reproduce the
excitation energy of the doublet and the near degeneracy within threshold
extrapolation and convergence uncertainties. However, the order of levels is
wrong. In contrast, the Julich'04 interaction \cite{2005PhRvC..72d4005H} gives
a significant splitting of the spin doublet in contradiction to experiment.

The energy splitting of the $5/2_{1}^{+}-3/2_{1}^{+}$ doublet states in
$_{\Lambda}^{9}$Be, which was considered to be dominantly composed of the $^{8}$Be$(2_{1}^+)\bigotimes\Lambda(s_{1/2})$ configuration, has been studied in \cite{2000PhRvL..85..270H} within a microscopic three-body model $2\alpha
+\Lambda$. The Pauli principle between two $\alpha$ clusters has been taken
into account by the orthogonality condition model. The main purpose of Ref.
\cite{2000PhRvL..85..270H} was to demonstrate how the splitting of the spin-doublet states in
$_{\Lambda}^{9}$Be is related to the underlying LS and antisymmetric LS forces
(ALS), which are different between one-boson-exchange models and quark models.
The quark model predicts that the ALS component of the LN interaction is so
strong as to substantially cancel the LS one, while the one-boson-exchange
models propose much smaller ALS and various strength of LS. The $\Lambda
\alpha$ interactions are derived by folding the $\Lambda N$ interaction into
the density of the $\alpha$ cluster. The authors introduced a phenomenological
$\Lambda NN$ three-body force, folding of which leads to both $\Lambda
\alpha\alpha$ and $\Lambda\alpha$ potentials. All the available Nijmegen
one-boson-exchange model $\Lambda N$ interactions lead to a wide range of
splittings of 0.08-0.20 MeV in $_{\Lambda}^{9}$Be. At the same time,
quark-model $\Lambda N$ interactions, which have generally large ALS force,
gives a half of the smallest one-boson-exchange model prediction for the
splitting. These data are compatible with the experimental results reported in
\cite{1988PhRvC..38..854Y}.

Based on the Faddeev methodology calculations of $2\alpha+\Lambda$ system, which used two-cluster resonating-group method kernels, have been performed in
Ref. \cite{2004PhRvC..70b4002F}. The method, which was used in
\cite{2004PhRvC..70b4002F}, is equivalent to the pairwise orthogonality
condition model of three-cluster systems, interacting via two-cluster RGM
kernels. The three-range Minnesota force, which describes the $\alpha\alpha$
phase shifts, has been chosen as an effective two-nucleon interaction. A
simple two-range Gaussian potential for each spin-singlet and spin-triplet
state, generated from the phase-shift behavior of the quark-model
hyperon-nucleon interaction, has been used as a $\Lambda N$ force for
$\Lambda\alpha$ interaction. To solve the Faddeev equation, the authors
discretized the continuous momentum variable for the Jacobi coordinate
vectors. The authors stated that the $L^\pi=0^{+}$ ground state and the $L^\pi=2^{+}$ excited state of $_{\Lambda}^{9}$Be are well described by the contracted $2\alpha$ cluster structure with a weakly coupled $\Lambda$-particle in the dominant $s$-wave component. However,the energy gain for $_{\Lambda}^{9}$Be due to partial waves higher than the s-wave is claimed to be about 1.2 MeV, because oscillatory behavior of the $\alpha\alpha$ relative wave functions needs more partial waves with a larger energy gain.

In the present paper the structure of bound and resonance states in $^9_\Lambda$Be hypernucleus for the states $1/2\leq J\leq7/2$ of positive and negative parity is investigated with special emphasis on the impact of cluster polarization on the spectrum of the $^9_\Lambda$Be and elements of scattering matrix. The Pauli exclusion principle between $\alpha$-clusters is taken into account completely.
We employ an effective $\Lambda N$ single-channel interaction simulating the basic features of the Nijmegen meson-theoretical models NSC97f \cite{2000PhRvL..85..270H}, in which a cut-off parameter $k_F$ was adopted to reproduce the energy of the ground state of $^9_\Lambda$Be with respect to $2\alpha+\Lambda$ threshold.
As a $NN$ interaction the modified  Hasegawa-Nagata potential is chosen with the Majorana parameter being adjusted to give the experimentally observed energy of $^9$Be nucleus.
Parameters of resonance states are determined from analysis of the energy dependence of two-channel $S$-matrix for $\alpha-_{\Lambda}^{5}$He and $\Lambda-^{8}$Be scattering, provided that $_{\Lambda}^{5}$He and $^{8}$Be subsystems being in their ground states in entrance and exit channels.

The paper is organized as follows. Formulation of a microscopic three-cluster model used for the investigation of the $^9_\Lambda$Be hypernucleus is given in Section \ref{sec:model}. In Section \ref{sec:results} we analyze how the spectrum of bound and resonance states of $^9_\Lambda$Be depends on the polarization of two-cluster subsystems $^5_\Lambda$He and $^8$Be. The nature of the obtained resonance states in $^9_\Lambda$Be is also discussed in Section \ref{sec:results}. Conclusions are made in Section \ref{sec:concl}.

\section{Formulation of the model}
\label{sec:model}

Let us consider a microscopic Hamiltonian for a system consisting of 8
nucleons (two alpha-particles) and a $\Lambda$ hyperon:%
\begin{eqnarray}
\widehat{H} &=&-\frac{\hbar^{2}}{2m}\sum_{i=1}^{8}\frac{\partial^{2}}%
{\partial\mathbf{r}_{i}^{2}}-\frac{\hbar^{2}}{2m_{\Lambda}}\frac{\partial^{2}%
}{\partial\mathbf{r}_{\Lambda}^{2}}\label{eq:001}\\
&+& \sum_{i<j}^{8}V_{NN}\left(  \mathbf{r}%
_{i}-\mathbf{r}_{j}\right)  +\sum_{i=1}^{8}V_{N\Lambda}\left(  \mathbf{r}%
_{i}-\mathbf{r}_{\Lambda}\right) \nonumber %
\end{eqnarray}
where $m=\left(  938.272+939.565\right)  /2=$938.919 MeV/c$^{2}$ is  a nucleon mass and $m_{\Lambda}=$1115.683(6) MeV/c$^{2}$ is a mass of the $\Lambda$
hyperon. It is more expedient to use the mass of a nucleon $m$ as a unit mass
and than the mass of the hyperon $\overline{m}_{\Lambda}=m_{\Lambda}/m=$1.18826.
It is assumed that coordinates of nucleons and a coordinate of the hyperon
are determined in the center-of-mass system, and thus center of mass
motion is eliminated from the Hamiltonian.

The total Hamiltonian can be separated into nuclear and hypernuclear parts:
\begin{eqnarray}
\widehat{H}  &  =&\widehat{H}_{NN}+\widehat{H}_{N\Lambda},\nonumber\\
\widehat{H}_{NN}  &  =&-\frac{\hbar^{2}}{2m}\sum_{i=1}^{8}\frac{\partial^{2}%
}{\partial\mathbf{r}_{i}^{2}}+\sum_{i<j}^{8}V_{NN}\left(  \mathbf{r}%
_{i}-\mathbf{r}_{j}\right)  ,\label{eq:002A}\\
\widehat{H}_{N\Lambda}  &  =&-\frac{\hbar^{2}}{2m_{\Lambda}}\frac{\partial^{2}%
}{\partial\mathbf{r}_{\Lambda}^{2}}+\sum_{i=1}^{8}V_{N\Lambda}\left(
\mathbf{r}_{i}-\mathbf{r}_{\Lambda}\right). \label{eq:002B}%
\end{eqnarray}

Eigenfunctions of Hamiltonian (\ref{eq:001}) characterized with the total angular momentum J and energy E of the relative motion of the cluster will be sought in the form:
\begin{eqnarray}
\Psi_{EJ}  &  =&\sum_{L}\sum_{\alpha=1}^{2}\sum_{\lambda_{\alpha},l_{\alpha}%
}\widehat{\mathcal{A}}\left\{  \Phi_{1}\left(  ^{4}He\right)  \Phi_{2}\left(
^{4}He\right)  \Phi_{3}(\Lambda)  \right. \label{eq:010}\\
&  \times & \left. f_{\lambda_{\alpha},l_{\alpha};L}^{\left(
E,J\right)  }\left(  x_{\alpha},y_{\alpha}\right) \left\{  Y_{\lambda_{\alpha}}\left(  \widehat{\mathbf{x}%
}_{\alpha}\right)  Y_{l_{\alpha}}\left(  \widehat{\mathbf{y}}_{\alpha}\right)
\right\}  _{L}\right\}  _{JM_{J}}.\nonumber
\end{eqnarray}
Here we involve two Faddeev amplitudes $f_{\lambda_{\alpha},l_{\alpha};L}^{\left(
E,J\right)  }\left(  x_{\alpha},y_{\alpha}\right)  $ which represent
dynamics in binary channels $^{5}_\Lambda$He+$\alpha$ ($\alpha$=1) and $^{8}$Be+$\Lambda$
($\alpha$=2).
The Jacobi vector $\mathbf{x}_{1}$ (=$x_{1}\cdot
\widehat{\mathbf{x}}_{1}$) determines distance between an alpha particle and a $\Lambda$-hyperon:
\begin{equation}
\mathbf{x}_1=\sqrt{\frac{4\,\overline{m}_{\Lambda}}{\overline{m}_{\Lambda}+4}
}\left[\mathbf{r}_{\Lambda}-\frac{1}{4}\sum_{i=1}^4\mathbf{r}_{i}\right], \label{eq:0111}
\end{equation}
while the Jacobi vector $\mathbf{y}_{1}$ is the distance between an alpha particle and $^5_\Lambda$He binary subsystem:
\begin{equation}
\mathbf{y}_1=\sqrt{\frac{4(\overline{m}_{\Lambda}+4)}{\overline{m}_{\Lambda}+8}
}\left[\frac{1}{4}\sum_{i=5}^8\mathbf{r}_{i}-\frac{1}{\overline{m}_{\Lambda}+4}\left(\mathbf{r}_{\Lambda}+\sum_{i=1}^4\mathbf{r}_{i}\right)\right] \label{eq:0121}
\end{equation}
 The second tree of Jacobi coordinates involves vector $\mathbf{x}_{2}$, which describes the relative distance between two alpha particles,
\begin{equation}
\mathbf{x}_2=\sqrt{2}\left[\frac{1}{4}%
\sum_{i=1}^4\mathbf{r}_{i}-\frac{1}{4}\sum_{j=5}^8\mathbf{r}%
_{j}\right]  \label{eq:011}%
\end{equation}
and vector $\mathbf{y}_{2}$, which determines position of the $\Lambda$-hyperon relative to the $^{8}$Be:
\begin{equation}
\mathbf{y}_2=\sqrt{\frac{8\,\overline{m}_{\Lambda}}{\overline{m}_{\Lambda}+8}
}\left[\mathbf{r}_{\Lambda}-\frac{1}{8}\sum_{i=1}^8\mathbf{r}_{i}\right]
\label{eq:012}%
\end{equation}

It is worthwhile underlining that the antisymmetrization operator $\widehat
{\mathcal{A}}$ in (\ref{eq:010}) permutes coordinates of nucleons only. It
does not involve a hyperon. Due to this fact, in the second Jacobi tree adopted for describing the channel $^{8}$Be+$\Lambda$
we have got a folding type of function $\Psi_{EJ}$ in (\ref{eq:010})  with
the wave function of  $^{8}$Be being antisymmetric. In the first Jacobi tree
associated with the channels $_{\Lambda}^{5}$He+$\alpha$ the
antisymmetrization operator $\widehat{\mathcal{A}}$ invokes the exchange of
nucleons between $_{\Lambda}^{5}$He and an alpha particle and thus makes
antisymmetric a wave function of the compound system $^9_\Lambda$Be.

Equation (\ref{eq:010}) represents the wave function in the $LS$ coupling scheme. Partial orbital momentum $\lambda_{\alpha}$ indicates an internal orbital momentum of $^{5}_\Lambda$He\ ($\alpha$=1) or $^{8}$Be
($\alpha$=2), while orbital momentum $l_{\alpha}$ describes rotation of an alpha particle around $^{5}_\Lambda$He\ ($\alpha$=1) or rotation of the $\Lambda$-hyperon around $^{8}$Be ($\alpha$=2). The total orbital momentum $L$ is a vector sum
of partial orbital momenta: $\overrightarrow{L}=\overrightarrow{\lambda }_{\alpha}+\overrightarrow{l}_{\alpha}$. Since the spins of alpha-clusters are equal to zero, the total spin of the hypernucleus $^9_\Lambda$Be is determined by the spin of the $\Lambda$-hyperon and equals 1/2.  Thus with a given value of the
total angular momentum $J$ the total orbital momentum $L$ can have two values
$L=J-1/2$ and $L=J+1/2$. It is true for all values of $J$ and parity $\pi$ except when $J^{\pi}=1/2^{-}$ where the total orbital momentum has only one value $L=1$.

Faddeev three-cluster amplitudes $f_{\lambda_{\alpha},l_{\alpha};L}^{\left(
E,J\right)  }\left(  x_{\alpha},y_{\alpha}\right)  $  are the solutions of an infinite set of integro-differential equations resulting from Schr\"{o}dinger equation for the wave function
(\ref{eq:010}) with the Hamiltonian (\ref{eq:001}):
\begin{eqnarray}
&&  \left[  \widehat{T}_{x_{\alpha},\lambda_{\alpha}}+\widehat{T}_{y_{\alpha
},l_{\alpha}}-E\right]  f_{\lambda_{\alpha},l_{\alpha};L}^{\left(
E,J\right)  }\left(  x_{\alpha},y_{\alpha}\right)    \nonumber\\
& +& \sum_{\beta=1}^{2}%
\sum_{\lambda_{\beta},l_{\beta}}\int_{0}^{\infty}\int_{0}^{\infty}%
d\widetilde{x}_{\beta}\widetilde{x}_{\beta}^{2}d\widetilde{y}_{\beta
}\widetilde{y}_{\beta}^{2}
\mathcal{V}_{\lambda_{\alpha},l_{\alpha};\lambda_{\beta},l_{\beta}%
}^{\left(  L\right)  }\left(  x_{\alpha},y_{\alpha};\widetilde{x}_{\beta
},\widetilde{y}_{\beta}\right)  \cdot f_{\lambda_{\beta},l_{\beta};L}^{\left(  E,J\right)
}\left(  \widetilde{x}_{\beta},\widetilde{y}_{\beta}\right)
 \label{eq:102} \\
&  = & E\sum_{\beta=1}^{2}\sum_{\lambda_{\beta},l_{\beta}}\int_{0}^{\infty}%
\int_{0}^{\infty}d\widetilde{x}_{\beta}\widetilde{x}_{\beta}^{2}d\widetilde
{y}_{\beta}\widetilde{y}_{\beta}^{2}
\mathcal{N}_{\lambda_{\alpha},l_{\alpha};\lambda_{\beta},l_{\beta}%
}^{\left(  L\right)  }\left(  x_{\alpha},y_{\alpha};\widetilde{x}_{\beta
},\widetilde{y}_{\beta}\right)  \cdot f_{\lambda_{\beta},l_{\beta};L}^{\left(  E,J\right)
}\left(  \widetilde{x}_{\beta},\widetilde{y}_{\beta}\right)
,\nonumber
\end{eqnarray}
where%
\[
\widehat{T}_{z,l}=-\frac{\hbar^{2}}{2m}\left[  \frac{d^{2}}{dz^{2}}+\frac
{2}{z}\frac{d}{dz}-\frac{l\left(  l+1\right)  }{z^{2}}\right]
\]
is the kinetic energy operator associated with the Jacobi vector
$\mathbf{z}=\mathbf{x}_{\alpha}$ or $\mathbf{z}=\mathbf{y}_{\alpha}$,  $\mathcal{N}_{\lambda_{\alpha},l_{\alpha};\lambda_{\beta},l_{\beta}}^{\left(L\right)}$ is the exchange part of the norm kernel, and $\mathcal{V}_{\lambda_{\alpha},l_{\alpha};\lambda_{\beta},l_{\beta}}^{\left(  L\right)}$ contains a direct and an exchange
part of the potential energy and an exchange term of the kinetic energy of the three-cluster system.

We can reduce a three-cluster problem to a many-channel two-body problem by expanding the wave function $f_{\lambda_{\alpha},l_{\alpha};L}^{\left(  E,J\right)
}\left(  x_{\alpha},y_{\alpha}\right) $ into the basis of eigenfunctions of the two-cluster Hamiltonian consisting of bound states $g_{\mathcal{E}^{\sigma}_\alpha\lambda_{\alpha}}\left(x_{\alpha}\right)  $ ($\sigma$= 1, 2, \ldots) and continuous spectrum states $g_{\mathcal{E}_\alpha\lambda_{\alpha}}\left(
x_{\alpha}\right) :$
\begin{eqnarray}
f_{\lambda_{\alpha},l_{\alpha};L}^{\left(  E,J\right)
}\left(  x_{\alpha
},y_{\alpha}\right)  &=& \sum_{\sigma}g_{{\mathcal{E}^{\sigma}_\alpha\lambda_{\alpha}}}
\left(  x_{\alpha}\right)  \cdot\phi_{E-\mathcal{E}_\alpha^{\sigma},\,l_{\alpha}}\left(  y_{\alpha}\right)  \label{eq:115} \\
&+&\int d\mathcal{E}_\alpha g_{\mathcal{E}_\alpha\lambda_{\alpha}}\left(  x_{\alpha}\right)
\phi_{E-\mathcal{E}_\alpha,\,l_{\alpha}}\left( y_{\alpha}\right) . \nonumber%
\end{eqnarray}
Functions $g_{\mathcal{E}^{\sigma}_\alpha\lambda_{\alpha}}\left(x_{\alpha}\right) $ and $g_{\mathcal{E}_\alpha\lambda_{\alpha}}\left(x_{\alpha}\right)$  satisfy the two-cluster Schr\"{o}dinger equation:
\begin{eqnarray}
& & \left[  \widehat{T}_{x_{\alpha},\lambda_{\alpha}}-\mathcal{E}_\alpha\right]  g_{\mathcal{E}_\alpha\lambda
_{\alpha}}\left(  x_{\alpha}\right) \label{eq:110}\\
& +&\int
_{0}^{\infty}d\widetilde{x}_{\alpha}\widetilde{x}_{\alpha}^{2}\cdot
\mathcal{V}^{\left(  \lambda_{\alpha}\right)  }\left(  x_{\alpha}%
;\widetilde{x}_{\alpha}\right)  \cdot g_{\mathcal{E}_\alpha\lambda_{\alpha}}\left(  \widetilde{x}_{\alpha}\right)  \nonumber \\
&  =&\mathcal{E}_\alpha\int_{0}^{\infty}d\widetilde{x}_{\alpha}\widetilde{x}_{\alpha}^{2}%
\cdot\mathcal{N}^{\left(  \lambda_{\alpha}\right)  }\left(  x_{\alpha
};\widetilde{x}_{\alpha}\right)  \cdot g_{\mathcal{E}_\alpha\lambda_{\alpha}}\left(  \widetilde{x}_{\alpha}\right)  .\nonumber
\end{eqnarray}
Functions $\phi_{E-\mathcal{E}^{\sigma}_\alpha,\,l_{\alpha}}\left( y_{\alpha}\right)$  and $\phi_{E-\mathcal{E}_\alpha,\,l_{\alpha}}\left( y_{\alpha}\right)$ describe scattering of the third cluster on a two-cluster bound state with energy  $\mathcal{E}^{\sigma}_\alpha$ or a continuum spectrum state with energy $\mathcal{E}_\alpha$, correspondingly. Quantities $\mathcal{N}^{\left( \lambda_{\alpha}\right)  }\left(  x_{\alpha };\widetilde{x}_{\alpha}\right)$ and $\mathcal{V}^{\left(  \lambda_{\alpha}\right)  }\left(  x_{\alpha};\widetilde{x}_{\alpha}\right)$ have the same meaning as the similar quantities in Eq. (\ref{eq:102}), but for two-cluster systems.

So, first we solve two-cluster equation (\ref{eq:110}) and then use the eigenfunctions of the two-cluster Hamiltonian to find wave functions $\phi_{E-\mathcal{E}^{\sigma}_\alpha,\,l_{\alpha}}\left( y_{\alpha}\right)$  and $\phi_{E-\mathcal{E}_\alpha,\,l_{\alpha}}\left( y_{\alpha}\right)$. Finally, we get
three-cluster wave function $f_{\lambda_{\alpha},l_{\alpha};L}^{\left(  E,J\right)
}\left(  x_{\alpha},y_{\alpha}\right)$.

In practical calculations the integral part of  the expansion (\ref{eq:115}) is substituted with the sum over the finite number of the discretized states in two-cluster continuum. The more terms in this sum are, the better cluster polarization is taken into account.

As in \cite{2017NuPhA.958...78L}, we use a finite number of square-integrable Gaussian functions to expand two-cluster wave function $g_{\mathcal{E}%
\lambda_{\alpha}}\left(  x_{\alpha}\right):$
\begin{equation}
g_{\mathcal{E}\lambda_{\alpha}}\left(  x_{\alpha
}\right)  =\sum_{\nu=1}^{N_{G}^{max}}D_{\nu}^{\left(  \mathcal{E}\lambda_{\alpha
}\right)  }G_{\lambda_{\alpha}}\left(  x_{\alpha},b_{\nu}\right)  ,
\label{eq:117}%
\end{equation}
where
\begin{eqnarray}
G_{\lambda}\left(  \mathbf{x},b_{\nu}\right)  &= &
\sqrt{\frac{2}{b_{\nu}^{3}\Gamma\left(  \lambda_{\alpha}+3/2\right)  }}\rho
^{\lambda_{\alpha}}\exp\left\{  -\frac{1}{2}\rho^{2}\right\},  \label{eq:118} \\
&&\left(
\rho=\frac{x}{b_{\nu}}\right)   \nonumber %
\end{eqnarray}
is a Gaussian function. Parameters $b_\nu$ are chosen so to minimize the ground state energies of the two-body subsystems.

To find wave functions $\phi_{E-\mathcal{E},l_{\alpha}}\left(  y_{\alpha}\right) $ of the third cluster interacting with
the two-cluster subsystem numerated by index $\alpha$\ ($\alpha$=1,2), we expand them over the
oscillator basis
\begin{equation}
\phi_{E-\mathcal{E},l_{\alpha}}\left(  y_{\alpha}\right) =\sum_{n_{\alpha}%
=0}^{N_{0}-1}C_{n_{\alpha}}^{\left(E-\mathcal{E},\,l_{\alpha}\right)  }%
\psi_{n_{\alpha},l_{\alpha}}\left(  y_{\alpha},b\right)  , \label{eq:120}%
\end{equation}
where%
\begin{eqnarray}
\psi_{n_{\alpha},l_{\alpha}}\left( y_{\alpha},b\right)   &  =& \left(
-1\right)  ^{n_{\alpha}}\mathcal{N}_{n_{\alpha}l_{\alpha}}%
{\tilde \rho}^{l_{\alpha}}e^{-\frac{1}{2}{\tilde \rho}^{2}}L_{n_{\alpha}}^{l_{\alpha}
+1/2}\left({\tilde \rho}^{2}\right)  ,\quad\label{eq:121}\\
{\tilde \rho}  = \frac{y_{\alpha}}{b}, && \quad\mathcal{N}_{n_{\alpha}l_{\alpha}}%
=\sqrt{\frac{2\Gamma\left(  n_{\alpha}+1\right)  }{b^{3}~\Gamma\left(  n_{\alpha
}+l_{\alpha}+3/2\right)  }}\nonumber
\end{eqnarray}
is an oscillator function and $b$ is the oscillator length.

In eq.
(\ref{eq:120}), a finite number of oscillator functions appears. But
in fact, this expansion involves an infinite number of functions, since
we know an asymptotic behavior of wave function $\phi_{E-\mathcal{E},\,l_{\alpha}}\left(  y_{\alpha}\right)$ in coordinate space and expansion coefficients
$C_{n_{\alpha}}^{\left(E-\mathcal{E},\,l_{\alpha}\right)  }$ in oscillator space.

At large distances $x_1\ll y_1$ between  an $\alpha$-particle and a $^{5}_\Lambda$He subsystem being in the state with energy $\mathcal{E}_1$  wave function $\phi_{E-\mathcal{E}_1,\,l_{1}}\left(  y_{1}\right)$ has the following form:
\begin{equation}
\phi_{E-\mathcal{E}_1,l_{1}}\left(  y_{1}\right)  \approx   \delta_{c_{0},c}
\psi_{l_{1}}^{\left(  -\right)  }\left(  k_{1}y_{1}
;\eta_{1}\right)  -S_{c_{0},c}\psi_{l_{1}}^{\left(  +\right)
}\left(  k_{1}y_{1};\eta_{1}\right) , \label{eq:125}%
\end{equation}
where $S_{c_{0},c}$ is the scattering matrix, index $c$ numerates an exit channel $c=\left\{  \mathcal{E}_{1
}\lambda_{1}l_{1}\right\}  $ and $c_{0}$ indicates the entrance
channel, $\psi_{l_{1}}^{\left(-\right)}$ ($\psi_{l_{1}}^{\left(+\right)}$) is the incoming
(outgoing) Coulomb wave, and $\eta_{1}$ is Sommerfeld parameter.
Determination of the incoming and outgoing Coulomb wave functions can be
found, for instance, in Ref. \cite{1983RvMP...55..155B}.

The asymptotic behaviour of the wave function $\phi_{E-\mathcal{E}_2,\,l_{2}}\left(  y_{2}\right)$ describing scattering of a $\Lambda$-hyperon on $^8$Be subsystem being in the state with energy $\mathcal{E}_2$  is determined by a superposition of the Hankel functions $H_{l_{2}+1/2}^{\left(\pm\right)  }$, since $\Lambda$-hyperon does not have an electric charge:
\begin{equation}
\phi_{E-\mathcal{E}_2,l_{2}}\left( y_{2}\right)  \approx   \delta_{c_{0},c}
H_{l_{2}+1/2}^{\left(  -\right)  }\left(  k_{2}y_{2}\right)  -S_{c_{0},c}H_{l_{2}+1/2}^{\left(  +\right)
}\left(  k_{2}y_{2}\right) , \label{eq:126}%
\end{equation}
Parameters
$k_{1,2}$ and $\eta_{1}$ in our case are defined as%
\begin{eqnarray*}
k_{1,2}  &  =& \sqrt{\frac{2m\left(  E-\mathcal{E}_{1,2}\right)  }{\hbar^{2}}},\\
\eta_{1}  &  =& \frac{Z^2  e^{2}%
}{\sqrt{2\left(  E-\mathcal{E}_{1,2}\right)  }}\sqrt{\frac{m}{\hbar^{2}}%
\frac{4\left(  4+\overline{m}_{\Lambda}\right)  }{\overline{m}_{\Lambda}+8}},
\end{eqnarray*}
where $Z=2$ is a charge of $\alpha$-cluster, $E$ is the total energy of three-cluster system ($E>\mathcal{E}_{1,2}$).

Having obtained all elements $S_{c,\widetilde{c}}$ of the scattering $S$
matrix, we make the following steps to extract important physical information.
First, we use\ one of the standard parametrizations\ of the $S$ matrix:%
\[
S_{c,\widetilde{c}}=\eta_{c,\widetilde{c}}\exp\left\{  2i\delta_{c,\widetilde
{c}}\right\}  .
\]
The diagonal values of the phase shifts $\delta_{c,c}$ and inelastic
parameters $\eta_{c,c}$ are analyzed to study elastic and inelastic processes
in a many-channel system. Second, the $S$ matrix $\left\Vert S_{c,\widetilde
{c}}\right\Vert $ is reduced to the diagonal form or to the representation of
the uncoupled channels. In this representation we obtain a set of eigenphase
shifts $\delta_{\alpha}$ which provides us with additional information on the
processes under consideration. The eigenphase shifts $\delta_{\alpha}$ also
allow us to determine the total and partial widths of resonance states in the
compound system. The details of this scheme and its justifications can be found in
Ref. \cite{2007JPhG...34.1955B}.

For numerical investigations of the three-cluster system $_{\Lambda}^{9}$Be,
we have to use a finite basis of Gaussian and oscillator functions.
As was pointed out above, $N_{G}^{max}$
Gaussian functions give us the same number of eigenstates and corresponding
eigenfunctions of a two-cluster Hamiltonian. To study effects of cluster
polarization, we will involve different numbers of these eigenstates, their
actual number we denote as $N_{G}$ (1$\leq N_{G}\leq
N_{G}^{max}$).

Index $N_{f}$ numerates a cumulative number of a basis function for
the $\sigma_\alpha$th\ eigenstate of the $\alpha$th ($\alpha$=1, 2)
two-cluster subsystem with energy $\mathcal{E}_{\sigma_{\alpha}}$ (1$\leq\sigma_{\alpha}\leq N_{G}$)
and for $n_\alpha$ oscillator quanta (0$\leq n_\alpha\leq N_{O}-1$):
\begin{eqnarray}
N_{f} &=&1+n_\alpha+\left(\sigma_{\alpha}-1\right)N_{O}+\left(\alpha-1\right)  N_{G}N_{O},\label{nf}\\
& & 1\leq N_f\leq 2N_GN_O. \nonumber
\end{eqnarray}
Thus, in our calculations we deal with the $2N_G$-channel system for the case of two coupled binary configurations. In an isolated configuration approximation the number of channels equals $N_G$.

Traditionally, the spectrum of the bound states is obtained with the
diagonalization  of the three-cluster Hamiltonian with $N_{f}$\ basis functions.
The diagonalization also reveals a large numbers of pseudo-bound states which are specific combinations of scattering states. In what follows, we are going to
study both bound and pseudo-bound states.

To obtain scattering wave functions and elements of scattering $S$-matrix, we
will solve a system of nonhomogeneous algebraic equations for the expansion coefficients.
For scattering states, the number of oscillator functions $N_{O}$ determines a border between an internal and asymptotic regions. It is obvious that the number of channels and the number of basis functions can be different for calculating bound  and scattering states.

We usually perform two type of calculations of the bound and scattering states. In the first type of calculations we take into account
polarizability of clusters involving maximal number of eigenstates of the
two-cluster Hamiltonians, i.e. $1\leq\sigma_{\alpha}\leq N_{G}^{\max}$. This approach will be called the approach with ``soft'' clusters. The second type of
calculations discards polarizability of clusters by involving only one
eigenfunction ($\sigma_{\alpha}=1$) of two-cluster Hamiltonians. It is
obviously that in such a case clusters do not change their size and shape, and
thus we call it the approach with ``rigid'' clusters.

\section{Results and Discussions}
\label{sec:results}
In the present paper we restrict the discussion to the case of zero
orbital momenta $\lambda_{\alpha}$ of binary subsystems $^5_\Lambda$He
and $^8$Be. Hence,
orbital momentum $l_{\alpha}$ describing relative motion of the binary
subsystems and the third cluster determines the total orbital momentum
$L=l_{\alpha}$ and parity. Consequently, a state with a given total
angular momentum and parity $J^\pi$ corresponds to the only value of $L$. Namely, positive parity states are characterized by even values of $L$, while negative parity states correspond to the odd values of $L$. $J^\pi=1/2^+$ states have zero total orbital momentum.

In many publications devoted to the hypernucleus $_{\Lambda}^{9}$Be, the bound and
resonance states were marked by the total orbital momentum $L$ and parity
$\pi$, which was explained by a small contribution of the spin-orbit
components of a  Lambda-nucleon interaction. To make a bridge between this
notation and ours, we will mark the states of $_{\Lambda}^{9}$Be with three
quantum numbers $J$, $\pi$ and $L$ in the following way $J^{\pi}$($L$).

In the asymptotic region we have taken into account two channels describing
the scattering of an $\alpha$-particle on $^5_\Lambda$He subsystem and
a $\Lambda$-hyperon on $^8$Be subsystem, provided that both subsystems
are in their ground states. In the internal region three more excited
states for each subsystem have also been considered. Such an approximation
allows for polarization of two-cluster subsystems due to the interaction
with the third cluster at small distances between clusters, but at large
distances binary subsystems
can be only in their ground states.

\subsection{Input parameters}

First of all we need to select values for the input parameters. We perform and present two sets of calculations by using two different sets of
input parameters. These sets realize two criteria for selecting input
parameters which reproduce different observable quantities. We will denote two
sets of input parameters $P1$ and $P2$, correspondingly. We start with the
first set of the input parameters and in Subsections \ref{subsubsec:convergence}, \ref{subsubsec:convergence} the main results will be discussed  with the
$P1$. In Subsection \ref{subsubsec:P2}  we consider $P2$ set of input parameters.

\begin{figure}[ptbh]
\begin{center}
\includegraphics[width=\columnwidth]{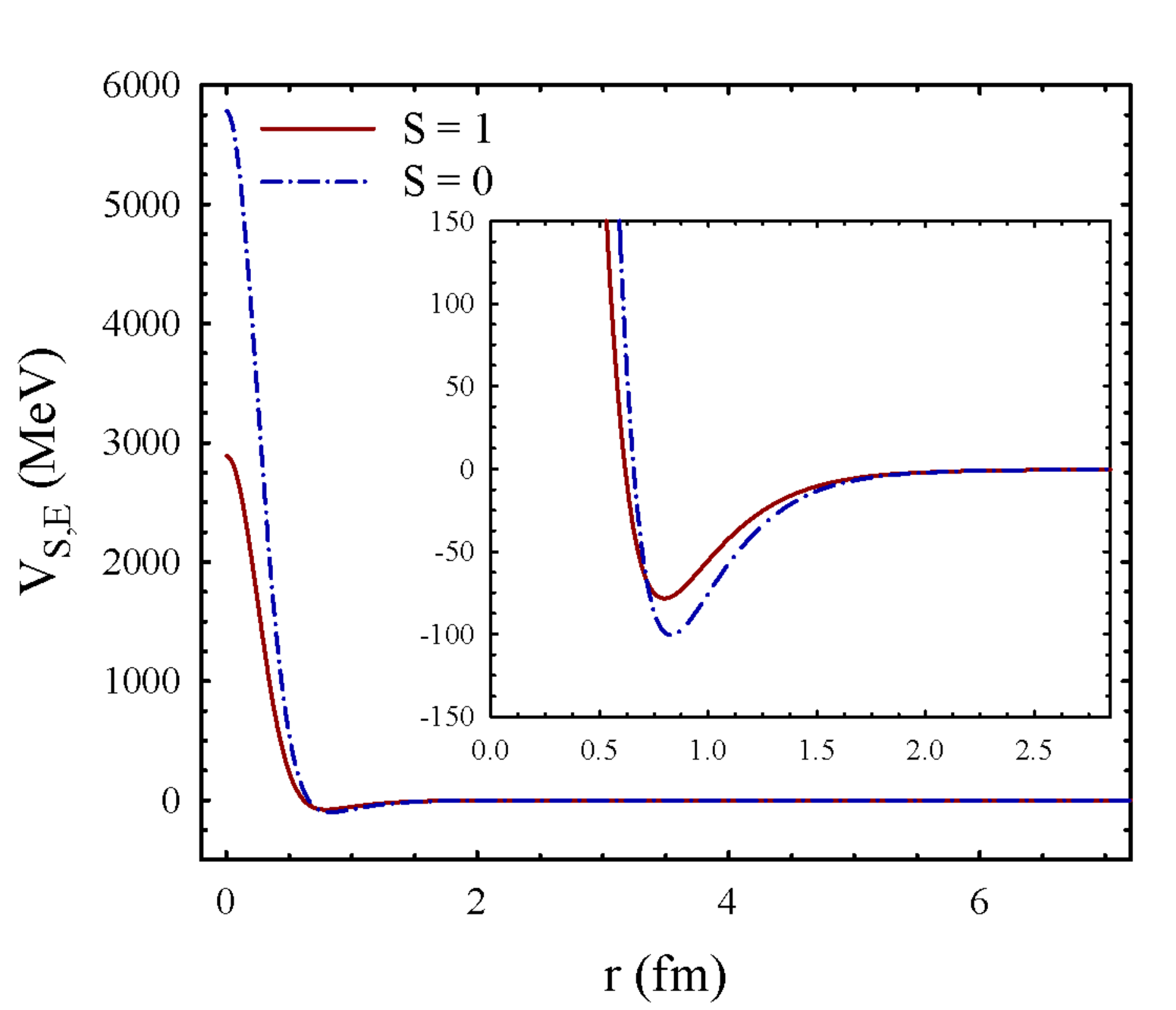}
\end{center}
\caption{The even componenents of the central part of the $N\Lambda $ potential.}
\label{Fig:HNPoten1Even}
\end{figure}

We involve the modified Hasegawa-Nagata (MHNP) \cite{potMHN1, potMHN2}
potential as a nucleon-nucleon interaction. We use the same nucleon-hyperon
potential as in Ref. \cite{2006PhRvC..74e4312H}. Parameters of the spin-orbit
interaction of the $\Lambda N$ interaction  are taken for the version NSC97f from
Ref. \cite{2000PhRvL..85..270H}.
In Fig. \ref{Fig:HNPoten1Even}\ we display the even components of the
nucleon-hyperon potential.

Traditionally we chose the oscillator length $b=1.317$ fm to minimize
the threshold energy of the two-cluster subsystems. The value $m=0.4389$ of
Majorana parameter of the MHNP is adjusted to reproduce the energy and
width of the ground state in $^{9}$Be nucleus relative to $\alpha+\alpha+n$
threshold. The standard value of the Majoran parameter is $m=0.4057$.
Cut-off parameter $k_{eff}=0.889$ fm$^{-1}$ of the $\Lambda N$-potential
is selected to reproduce the energy of the ground state of $^{9}_\Lambda$Be
nucleus with respect to $\alpha+\alpha+\Lambda$ threshold.

Parameters $b_{\nu}$ ($\nu$=1, 2, 3, 4) of
Gaussian functions are determined as $b_{\nu}=b_{0}q^{\nu-1}$. In $P1$ input parameter set, for
$^{5}_\Lambda$He we selected $b_{0}=0.7$ fm, $q=1.85$, and for $^{8}$Be
we took $b_{0}=1.15$ fm, $q=2.2$. Such values of the parameters minimize
the ground state energies of the $^5_\Lambda$He and $^{8}$Be nuclei.

We have also invoked for the calculations another set of the input parameters ($P2$), which gives the experimental value $-3.12$ MeV for the binding energy of the $^5_\Lambda$He. Namely, we used different values of $b_{0}=0.25$ fm, $q=1.85$ for the Gaussian functions describing the $^5_\Lambda$He subsystem and 
slightly adjusted cut-off parameter $k_{eff}=0.887$ fm$^{-1}$ of the $\Lambda N$-potential to reproduce the energy of the ground state of $^{9}_\Lambda$Be nucleus with respect to $\alpha+\alpha+\Lambda$ threshold. Oscillator length $b=1.317$ fm and the value $m=0.4389$ of Majorana parameter of the MHNP and parameters of the Gaussian functions, describing the $^8$Be subsystem, remained the same in $P2$ input parameter set.

Note, that with such values of oscillator length $b$ and the Majorana
parameter $m$ of the MHNP we obtain resonance states of $^{8}$Be which are
presented in Table \ref{Tab:8BeSpectr}. In Table \ref{Tab:8BeSpectr} we also show
the experimental parameters of resonance states in $^{8}$Be. As we see, the theoretical
results are in fairly good agreement with the available experimental data
presented in Ref. \cite{2004NuPhA.745..155T}. It was indicated in Ref.
\cite{2017PhRvC..96c4322V} that these input parameters are not optimal for
resonance structure of $^{8}$Be. The optimal values of $b$ and $m$ leads to
overbinding 
the compound nucleus $^{9}$Be.
This is a typical problem for many semi-realistic nucleon-nucleon potentials. Thus, in the present
calculations we use those values of $b$ and $m$ which provide the correct
position of the $^{9}$Be ground state with respect to the three-cluster
threshold $\alpha+\alpha+n$.%

\begin{table}[tbp] \centering
\caption{Spectrum of resonance states in $^8$Be calculated in a two-cluster model
and compared with experimental data \cite{2004NuPhA.745..155T}}%
\begin{tabular}
[c]{ccccc}\hline
& \multicolumn{2}{c}{Theory} & \multicolumn{2}{c}{Experiment}\\\hline
$J^{\pi}$ & $E$, MeV & $\Gamma$, MeV & $E$, MeV & $\Gamma$, MeV\\\hline
$0^{+}$ & 0.859 & 0.958 & 0.092 & 5.57 $\times$ 10$^{-6}$\\
$2^{+}$ & 4.138 & 4.809 & 3.12 & 1.513\\
$4^{+}$ & 14.461 & 6.386 & 11.44 & $\approx$3.500\\ %
\hline
\end{tabular}
\label{Tab:8BeSpectr}%
\end{table}%

\subsection{Spectrum of $^9_\Lambda$Be nucleus}

\subsubsection{Convergence of parameters of resonance states, $P1$ input parameters}
\label{subsubsec:convergence}

The spectrum of positive and negative parity  states for $1/2\leq J\leq 7/2$  ($0\leq L\leq 4 $) of
the $^9_\Lambda$Be nucleus calculated  within the AMGOB model  is presented in
Tables \ref{Tab:Spectrum9Be+Gob}, \ref{Tab:Spectrum9Be-Gob}.
Tables \ref{Tab:Spectrum9Be+Gob}, \ref{Tab:Spectrum9Be-Gob}  demonstrate that the
polarization of two-cluster subsystems plays an important role in formation of
bound and resonance states of the $^9_\Lambda$Be.
\begin{table}[tbph] \centering%
\caption{Spectrum of positive parity states of $^9_\Lambda$Be nucleus obtained
within the AMGOB model with $P1$ set of the input parameters taking into account cluster polarization (soft) or neglecting
it (rigid). $\Gamma$ is a total width of the resonance state,
$\Gamma_{1,2}$ are partial decay widths of the resonance via $^5_\Lambda$He$+\alpha$
and $^8$Be$+\Lambda$ channels, correspondingly. Energy is given in MeV, the
total and partial widths are in keV.}%
\begin{tabular}{ccccccccc}
\hline
 & \multicolumn{4}{c}{rigid} & \multicolumn{4}{c}{soft}  \\ \hline
$J^{\pi}(L)$ & $E$ & $\Gamma $ & $\Gamma_1$ & $\Gamma_2$ & $E$ & $\Gamma $  & $\Gamma_1$ & $\Gamma_2$\\ \hline
$1/2^{+}(0)$& -6.204 &  &   &  &-6.623 &  &  & \\ \hline
        &5.360 & 3170 & 170 & 3000   & 1.792 &  3.3 & 2.3 & 1 \\ 
                &  &  &       &      & 2.252 & 60.3 & 11.8 & 48.5\\ 
                &  &  &       &      & 2.733 &  384& 0.6 & 383.4\\ 
                &  &  &       &      & 3.396 &  349.2& 84.6& 264.6\\ 
                &  &  &       &      & 4.107 &  213.6 & 23.3 & 190.3 \\ 
                &  &  &       &      & 4.650 &  470.8 & 77.8 & 393\\ 
                &  &  &       &      & 5.083  &  10.8 & 8.4 & 2.4\\ \hline
$3/2^{+}(2)$&-3.297 & 35.8 & 35.8 &  & -3.543 &  5.5 & 5.5 & \\ 
                &  &  &       &      & 2.115  & 0.045 & 0.01 & 0.035\\ 
                &  &  &       &      & 2.850 &  0.027 & 0.01 & 0.017\\ 
                &  &  &       &      & 3.314  & 206.9 & 0.1 & 206.8\\ 
                &  &  &       &      & 3.886 & 8.3 & 2 & 6.3\\ 
                &  &  &       &      & 4.387  & 105.7 & 1.6 & 104.1\\ 
                &  &  &       &      & 5.212 & 26.2 & 4.3 & 21.9\\ 
                &  &  &       &      & 5.610  &  55.8 & 0.4 & 55.4\\ \hline
$5/2^{+}(2)$ &  -3.446 & 12 & 12 & & -3.748 &  0.75 &  0.75 & \\ 
                &   &  &  &      & 2.115  & 0.043 & 0.01 & 0.033\\ 
                &   &  &  &      & 2.849 &  0.041 & 0.037 & 0.004\\ 
                &   &  &  &      & 3.312  & 210.42 & 0.07 & 210.35\\ 
                &   &  &  &      & 3.885 & 8.3 & 2& 6.3\\ 
                &   &  &  &      & 4.386  & 107.8 & 1.3 & 106.5\\ 
                &   &  &  &      & 5.209 & 26.7 & 4.2 & 22.5\\ 
                &   &  &  &      & 5.630  &  57.64 &  0.17& 57.4\\ \hline
$7/2^{+}(4)$ &  4.726 & 2962.5 & 2962.4 & 0.1 & 2.610 &  0.0091 & 0.0088 & 0.0003\\ 
                &   &  &  &      & 3.643  & 0.002 & 0.0007 & 0.0013\\ 
                &   &  &  &      & 3.979 &  8.302 &  0.001 & 8.301\\ 
                &   &  &  &      & 4.500  & 2625.2 & 2625.2 & 0.01 \\ 
                &   &  &  &      & 5.139 & 1.6 & 0.09 & 1.51 \\ \hline
\end{tabular}%
\label{Tab:Spectrum9Be+Gob}
\end{table}
For the positive parity states, cluster polarization decreases energy of the bound and
resonance states. It also reduces significantly total width of the resonance
states. Cluster polarization, for example, decreases the energy of the
$_{\Lambda}^{9}$Be ground 1/2$^{+}$(0) state by 400 keV, and the energy of
the 3/2$_1^{+} (0)$ resonance state which determined in experiments as the second excited
state is reduced by 245 keV, while its total width decreases more than 6
times. Similar situation with the 5/2$_1^{+}$(2) resonance state which has
approximately the same energy. Cluster polarization shifts down the energy of
the resonance state by 302 keV and causes an eightfold decrease in its width.
Besides, allowing for polarization leads to the formation of a large number of
narrow resonance states in three-cluster continuum above the decay threshold
$^9_\Lambda$Be$\Rightarrow^8$Be$(0_2^+)+\Lambda$ located at 1.63 MeV above the three-cluster decay threshold.

\begin{table}[htbp] \centering
\caption{Spectrum of negative parity states of $^9_\Lambda$Be nucleus obtained
within the AMGOB model with $P1$ set of the input parameters taking into account cluster polarization (soft) or neglecting
 it (rigid). $\Gamma$ is a total width of the resonance state,
$\Gamma_{1,2}$ are partial decay widths of the resonance via $^5_\Lambda$He$+\alpha$
and $^8$Be$+\Lambda$ channels, correspondingly. Energy is given in MeV, the total and partial widths are in keV.}%
\begin{tabular}{ccccccccc}
\hline
 & \multicolumn{4}{c}{rigid} & \multicolumn{4}{c}{soft}  \\ \hline
$J^{\pi} (L)$ & $E$ & $\Gamma $ & $\Gamma_1$ & $\Gamma_2$ & $E$ & $\Gamma $  & $\Gamma_1$ & $\Gamma_2$ \\ \hline
$1/2^{-}(1)$&  0.723  & 4521 & 0.4 & 4520.6 & 0.724 & 4228.5 & 0.4 & 4228.1\\ 
                &         &      & & & 1.924 & 3.74 & 3.735 & 0.005\\ 
                &         &      & & & 2.519 & 30.1 & 21.1& 9\\ 
                &         &      & & & 3.364 & 259.3 & 72.5 & 186.8\\ 
                &         &      & && 4.350 & 250.8 & 0.1& 250.7\\ 
                &         &      & & & 5.739 & 148.4 & 1.2& 147.2\\ 
                &         &      & & & 5.906 & 91.3 & 2.9& 88.4\\ \hline
$3/2^{-}(1)$ &  0.723 & 4507.4 & 0.4& 4507 & 0.724 & 4213.7 & 0.5 & 4213.2\\ 
                &         &  & &     & 1.924 & 3.99 & 3.97  & 0.02\\ 
                &         & & &     & 2.519 & 31.1 & 22& 9.1\\ 
                &         & & &     & 3.368 & 258.6 & 72.3& 186.3\\ 
                &         & & &     & 4.354 & 252.5 & 0.1& 252.4\\ 
                &         & & &     & 5.733 & 154 & 10& 144\\ 
                &         & & &     & 5.907 & 94.8 & 2.6 & 92.2\\ \hline
$5/2^{-}(3)$ &         &      & & & 2.345 & 0.132 & 0.003& 0.129\\ 
                &         & & &     & 3.227 & 0.1157 & 0.0007& 0.115\\ 
                &         & & &     & 4.117 & 54.1 & 0.6& 53.5\\ 
                &         & & &     & 4.403 & 1.5 & 0.4& 1.1\\ 
                &         & & &     & 5.133 & 42.5 & 11.4& 31.1\\ 
                &         & & &     & 5.866 & 2.3 & 1.1& 1.2\\ \hline
$7/2^{-}(3)$ &         &      & &      & 2.345 & 0.145 & 0.001& 0.144\\ 
                &         &  & &    & 3.227 & 0.197 & 0.001& 0.197\\ 
                &         & & &     & 4.117 & 54.3 & 0.6& 53.7\\ 
                &         & & &     & 4.403 & 1.4 & 0.4& 1\\ 
                &         & & &     & 5.132 & 41.7 & 11.5& 30.2\\ 
                &         & & &     & 5.865 & 2.1 & 1.1& 1\\ \hline
\end{tabular}%
\label{Tab:Spectrum9Be-Gob}
\end{table}%
For the negative parity states we observe from Table \ref{Tab:Spectrum9Be-Gob}
that the parameters of the lowest $1/2^{-}_1$(1) and $3/2^{-}_1$(1) resonance states
remain almost intact when cluster polarization is taken into account. The energy
 of this resonance is quite close to the energy 0.7 MeV where becomes possible
$^8$Be$(0_1^+)+\Lambda$ scattering. Since the latter channel is considered properly
already in the approximation of rigid two-cluster subsystems, allowing for cluster
polarization does not change a lot the lowest $1/2^{-}_1$(1) and $3/2^{-}_1$(1) states.
It might be well to point out that spectra of the $1/2^{-}$(1) and $3/2^{-}$(1) states are
almost degenerate, because they are characterized by the same orbital momentum $L=1$ and the spin-orbit interaction is small.

Analyzing partial widths of the resonance states tabulated in Tables
\ref{Tab:Spectrum9Be+Gob} and \ref{Tab:Spectrum9Be-Gob} we can observe
that the majority of the resonances decay via $^8$Be+$\Lambda$ channel.
There are not more than 2 resonances decaying via $^5_\Lambda$He+$\alpha$
channel for a given value of $J^\pi$.   In all the cases, except $5/2^-_6$(3) and $7/2^-_6$(3) resonances,
partial widths via different channels significantly differ from one another.
The above mentioned $5/2^-_6$(3) and $7/2^-_6$(3) states represent the only case when
the resonances decay with almost equal probability via both channels. It is
interesting to note also the principal change in partial widths of the lowest
$1/2^+_1$(0) resonance. Allowing for cluster polarization not only halves the energy
of this resonance state, but also changes its dominant decay channel.

It was pointed out in Ref. \cite{1985PThPS..81...42M} that the resonance
states 1/2$^{-}$(1) and 3/2$^{-}$(1)  in $_{\Lambda}^{9}$Be form the
"genuinely hypernuclear" rotational band as there is no such a band in the
nucleus $^{9}$Be. The Pauli principle forbids the valence nucleon to occupy
the lowest $p$ orbitals in $^{9}$Be, while there is no such a restriction for the
$\Lambda$\ particle in $_{\Lambda}^{9}$Be. In this respect it is interesting
to note that there are two types of the positive and negative parity states as
it follows from Tables \ref{Tab:Spectrum9Be+Gob} and \ref{Tab:Spectrum9Be-Gob}.
The first type comprises the resonance states with the dominant binary
structure  $\Lambda$ + $^{8}$Be. Such resonance states have the partial width
$\Gamma_{2}>\Gamma_{1}$. They are created mainly by rotation of the  $\Lambda
$\ particle arround  $^{8}$Be. And they are "genuinely hypernuclear" states.
In particular, we would like to note that our predictions concerning the energy and width of the lowest 1/2$^{-}_1$(1) and 3/2$^{-}_1$(1) resonance states
strongly correlate with the results obtained  in \cite{1985PThPS..81...42M}, since these states are located just near $\Lambda$ + $^{8}$Be and have quite large width.

The second type of states are represented by the resonance states with
 partial width $\Gamma_{1}>\Gamma_{2}$. These resonance states decay mainly via $\alpha$ + $_{\Lambda
}^{5}$He channel, where the  $\Lambda$\ particle forms a bound state with
an alpha particle. Rotation of the second alpha-particle around  $_{\Lambda
}^{5}$He creates a set of resonance states of the positive and negative parity.

In the energy region considered $E\leq 6$ MeV above $2\alpha+\Lambda$
threshold four open channels could play a part in the formation of resonance
states. They correspond to $\sigma=1,2$
eigenstates of the two-cluster subsystems calculated with four Gaussian
functions and listed in Table \ref{Tab:2CSystems}.
We can assume that appearance of narrow resonances in the case when cluster
polarization is taken into account could result from coupling the channels
belonging to the same cluster configuration: $^8$Be$(0_1^+)+\Lambda$ and
$^8$Be$(0_2^+)+\Lambda$ channels, $^5_\Lambda$He$(0_1^+)+\alpha$ and
$^5_\Lambda$He$(0_2^+)+\alpha$ channels. In particular, the first pair of
the channels could be more coupled due to proximity of the energies needed to make the channels open.

Referring to Table \ref{Tab:2CSystems} it will be observed that in our model
$^5_\Lambda$He subsystem is overbound by 0.9 MeV compared to its experimental
binding energy. However, this overbinding is much smaller than in other simple
model calculations based upon $\Lambda N$ potentials \cite{1995PhysRep...257..349G}.

In Table \ref{Tab:2CSystems} we also display the mass root-mean-square radius for all
bound and pseudo-bond states. As we see, the ground state of $_{\Lambda}^{5}%
$He is a compact two-cluster system, while the lowest pseudo-bound state,
representing the ground state of $^{8}$Be, is very dispersed with the large value
of $r_{m}=$5.71 fm. Dispersed are also states $\sigma=$2 and $\sigma=$3 of
$^{8}$Be and the state $\sigma=$2 of $_{\Lambda}^{5}$He, however with the
smaller values of $r_{m}$.%

\begin{table}[tbp] \centering
\caption{Energy and mass root-mean-square radius of the two-cluster bound and
pseudo-bound states with $P1$ set of the input parameters. The energy is measured from a two-cluster threshold
indicated in the first column.}
\begin{tabular}
[c]{cccccc}\hline
2C-system & Quantity & $\sigma=1$ & $\sigma=2$ & $\sigma=3$ & $\sigma
=4$\\\hline
$_{\Lambda}^{5}$He=$\alpha+\Lambda$ & $E$, MeV & -4.06 & 2.91 & 21.12 &
139.76\\
& $r_{m}$, fm & 1.71 & 3.25 & 2.20 & 1.48\\
& $\overline{r}_{\alpha}$, fm & 2.62 & 6.72 & 4.06 & 1.78\\
$^{8}$Be=$\alpha+\alpha$ & $E$, MeV & 0.70 & 1.63 & 9.11 & 53.03\\
& \multicolumn{1}{c}{$r_{m}$, fm} & 5.71 & 4.37 & 3.05 & 1.95\\
& $\overline{r}_{\alpha}$, fm & 15.16 & 7.07 & 2.34 & 1.39\\ \hline
\end{tabular}
\label{Tab:2CSystems}%
\end{table}%

Table \ref{Tab:2CSystems} shows also the average distances $\overline{r}_{\alpha}$ between
clusters. They are determined in the following way%
\[
\overline{r}_{\alpha}=b\sqrt{\left\langle g_{\mathcal{E}_{\sigma}%
,\lambda_{\alpha}}\left\vert x_{\alpha}^{2}\right\vert g_{\mathcal{E}_{\sigma
},\lambda_{\alpha}}\right\rangle /\mu_{\alpha}},
\]
where $\mu_{\alpha}$ is a reduced mass appearing in the definition of the
Jacobi vector $\mathbf{x}_{\alpha}$. The quantity $\overline{r}_{\alpha}$
gives
the most probable relative distance between
the interacting clusters. This quantity has been discussed in the literature. For
example, in Ref. \cite{2018PhRvC..97b4330K}\ devoted to the study of spectrum
of $p$-shell hypernuclei including $_{\Lambda}^{9}$Be within a microscopic
cluster model, the average distance between two alpha particles in the ground
state of $^{8}$Be were determined approximately. It was obtained
that $\overline{r}_{\alpha}$=5.99 fm which is smaller \ than our estimation
$\overline{r}_{\alpha}$=15.16 fm. We believe that such a large difference for
the average distance between alpha particles can be partially ascribed to
different nucleon-nucleon potentials involved in Ref.
\cite{2018PhRvC..97b4330K} and in our calculations. But the main difference is
related to the way how this quantity is determined. Note that detail investigation
of the average distance between two alpha particles in different states of $^{8}$Be
has been performed in Ref. \cite{2020arXiv200604525K} within the two-cluster resonating group method.

\subsubsection{Spectrum of the $3/2^+$(2) states, $P1$ input parameters}
\label{subsubsec:3/2+}

\begin{figure*}[ptbh]
\begin{center}
\includegraphics[width=\textwidth]{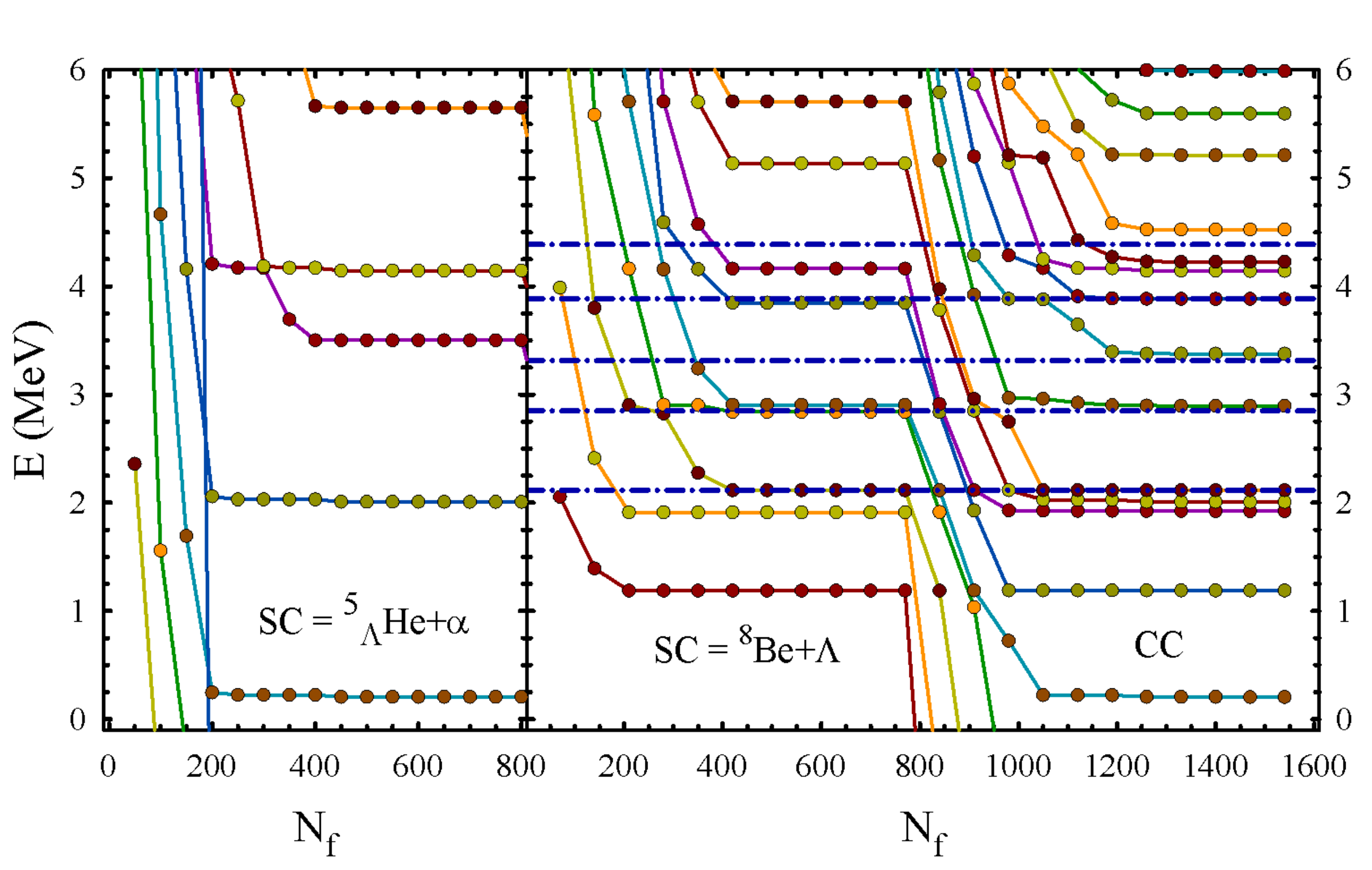}
\caption{Spectrum of eigenstates of the internal part of Hamiltonian for
3/2$^{+}$(2) states constructed in a single-configuration (SC) and
coupled-configuration (CC) approximations with $P1$ input paratemets. The 3/2$^{+}$(2) resonance states are
displayed by the dot-dashed line in the area of the CC approximation. Energy
is measured from the three-cluster decay threshold $\alpha+\alpha+\Lambda$.}%
\label{Fig:SpectrConv32PSCvsCC}%
\end{center}
\end{figure*}

In Fig. \ref{Fig:SpectrConv32PSCvsCC} we display the spectrum of
the Hamiltonian for the 3/2$^{+}$(2) states obtained with $P1$ input parameter set. The spectrum is obtained in a
single-configuration (SC) and coupled-configuration (CC) approximations.
Eigenvalues of the Hamiltonian are shown as a function of the number of
basis functions $N_{f}$ defined by Eq. (\ref{nf}). It is necessary to
recall that in the present calculations
the  single-configuration approximation involves four different channels
with a binary subsystem in the ground and three excited states.  200
oscillator functions are employed in each channel. Thus the region
1$\leq N_{f}\leq$800  in both panels of Fig.
\ref{Fig:SpectrConv32PSCvsCC} corresponds to the SC approximation, and the
region  801$\leq N_{f}\leq$1600 represents the CC approximation. The range
1$\leq N_{f}\leq$200 shows the spectrum of eigenstates created by basis
functions of the channel $_{\Lambda}^{5}$He$\left(  0_{1}^{+}\right)
$+$\alpha$ (left panel) and $^{8}$Be$\left(  0_{1}^{+}\right)  $+$\Lambda$
(right panel), the range 201$\leq N_{f}\leq$400 display effects
on the eigenspectrum of the second channel $_{\Lambda}^{5}$He$\left(  0_{2}%
^{+}\right)  $+$\alpha$ and $^{8}$Be$\left(  0_{2}^{+}\right)  $+$\Lambda$,
correspondingly, and so on.

One
can see that there are a large number of plateaus which appear in the SC
and CC approximations. The dot-dashed lines indicate the position of the 3/2$^{+}$(2)
resonance states, obtained by solving the dynamic equations
with proper boundary conditions.
Such a plateau can be a marker of a narrow resonance state in the system
under consideration. Besides, in many-channel systems the plateaus may appear
due to a weak coupling of channels.
It is worthwhile noticing that the second type of plateau may appear, for example,
when a weak spin-orbit interaction couples
states with different values of the total orbital momentum $L$ and/or
total spin $S$ ($LS$ coupling scheme).

Many plateaus which are observed in Fig. \ref{Fig:SpectrConv32PSCvsCC} in the
single-configuration approximations belong to the second type.
Indeed, if we have a closer look at the spectrum we can see that the
channel describing the interaction of the third cluster with the most compact
two-cluster subsystem ($_{\Lambda}^{5}$He$\left(0_{1}^{+}\right)$+$\alpha$ or
$^{8}$Be$\left(0_{1}^{+}\right)$+$\Lambda$) is dominant and noticeably reduces
the energy of the eigenstates. Second in importance to the eigenstates of the
$^9_\Lambda$Be hypernucleus are $_{\Lambda}^{5}$He$\left(0_{2}^{+}\right)+\alpha$
and $^{8}$Be$\left(0_{2}^{+}\right)+\Lambda$ channels.
All other channels are weakly coupled to the dominant channels
and thus contribute much less to the energy of the eigenstates shown in Fig.
\ref{Fig:SpectrConv32PSCvsCC}. The same is also valid for a number of plateaus
occurring in the CC area.
We can also see in Fig. \ref{Fig:SpectrConv32PSCvsCC} that in many cases the
energy of resonance states is close to the energy of plateau. The narrower is
a resonance state, the closer is its energy to the plateau energy.

\begin{table}[tbph] \centering%
\caption{Spectrum of $J^\pi(L)=3/2^+(2)$ states of $^9_\Lambda$Be nucleus obtained
within the AMGOB model with $P1$ set of the input parameters for different degree of polarization of two-cluster
subsystems. Energy is in MeV and width is in keV.}%
\begin{tabular}{cccccccc}
\hline
\multicolumn{2}{c}{$N_G=1$} &\multicolumn{2}{c}{$N_G=2$}& \multicolumn{2}{c}{$N_G=3$}& \multicolumn{2}{c}{$N_G=4$}  \\ \hline
$E$& $\Gamma $ & $E$ & $\Gamma $ & $E$ & $\Gamma $ & $E$ & $\Gamma $ \\ \hline
-3.297 & 35.8 &-3.430 & 13.9 & -3.539 & 5.6 &-3.543 &  5.5\\ 
        &      & 2.115 & 0.08 & 2.115 & 0.08& 2.115   & 0.04\\ 
       &      & 2.850 & 0.125 & 2.850 & 0.02& 2.850 &  0.03\\ 
         &      & 3.319 & 197.6& 3.314 & 206.9& 3.3136  & 206.9 \\ 
         &      & 3.887 & 9.4  & 3.886 & 8.3& 3.886  & 8.3 \\ 
         &      & 4.387 & 104.5 & 4.387 & 105.7& 4.387  & 105.7 \\ 
       &      & 5.217 & 32.1& 5.212 & 26.2& 5.212  & 26.2 \\ 
       &      & 5.643 & 68.6& 5.630 & 55.8& 5.630  &  55.8\\ \hline
\end{tabular}%
\label{Tab:Spectrum9Be3/2+Gob}
\end{table}

As Table \ref{Tab:Spectrum9Be3/2+Gob} suggests, the qualitative change
of the spectrum is caused by taking into account the  $0_2^+$ state in
$^{5}_\Lambda$He and $^{8}$Be subsystem. Allowing for higher excited states
leads
only to some slight alteration of energies and widths, but does not change
the  number of resonances. From this fact, we might reason that $0^+_2$
states of the binary subsystems plays a large part in the structure of
resonance states of $^9_\Lambda$Be nucleus.

In Table \ref{Tab:CoulombRS32P} we compare spectrum of the 3/2$^{+}$(2) resonance
states obtained with and without Coulomb forces. 200 oscillator functions are
used in both calculations.
\begin{table}[tbph] \centering
\caption{Effects of the Coulomb forces on the energy and width of the 3/2$^{+}$(2) resonance states studied with $P1$ set of the input parameters.}%
\begin{tabular}{cccc}
\hline
\multicolumn{2}{c}{with Coulomb} & \multicolumn{2}{c}{without Coulomb}  \\ \hline
$E$, MeV & $\Gamma$, keV & $E$, MeV & $\Gamma $, keV\\ \hline
-3.543 & 5.5 & -5.103 & 0  \\ 
2.115 & 0.04 & 1.129 & 1.122\\ 
2.850 & 0.03 & 1.865 & 0.107\\ 
3.314 & 206.9 & 2.764 & 29.034\\ 
3.886 & 8.3 & 3.126 & 64.264\\ \hline
\end{tabular}
\label{Tab:CoulombRS32P}
\end{table}
The results presented in Table \ref{Tab:CoulombRS32P} allow us to reveal
effects of the Coulomb forces \ on energy and width of the resonance states in
$_{\Lambda}^{9}$Be, and to understand peculiarities of the
present model.

We can conclude from Table \ref{Tab:CoulombRS32P} that without the Coulomb forces
energy of all the resonance states is decreased by approximately 1 MeV, and the
lowest $3/2_1^+$(2) resonance state with $E=-3.542$ MeV below the three-cluster threshold
is transformed into a bound state. It is interesting to note that the Coulomb
interaction makes $3/2_4^+$(2) resonance state
wider and other  resonance states narrower.

\begin{figure}[hptbh]
\begin{center}
\includegraphics[width=0.71\textwidth]{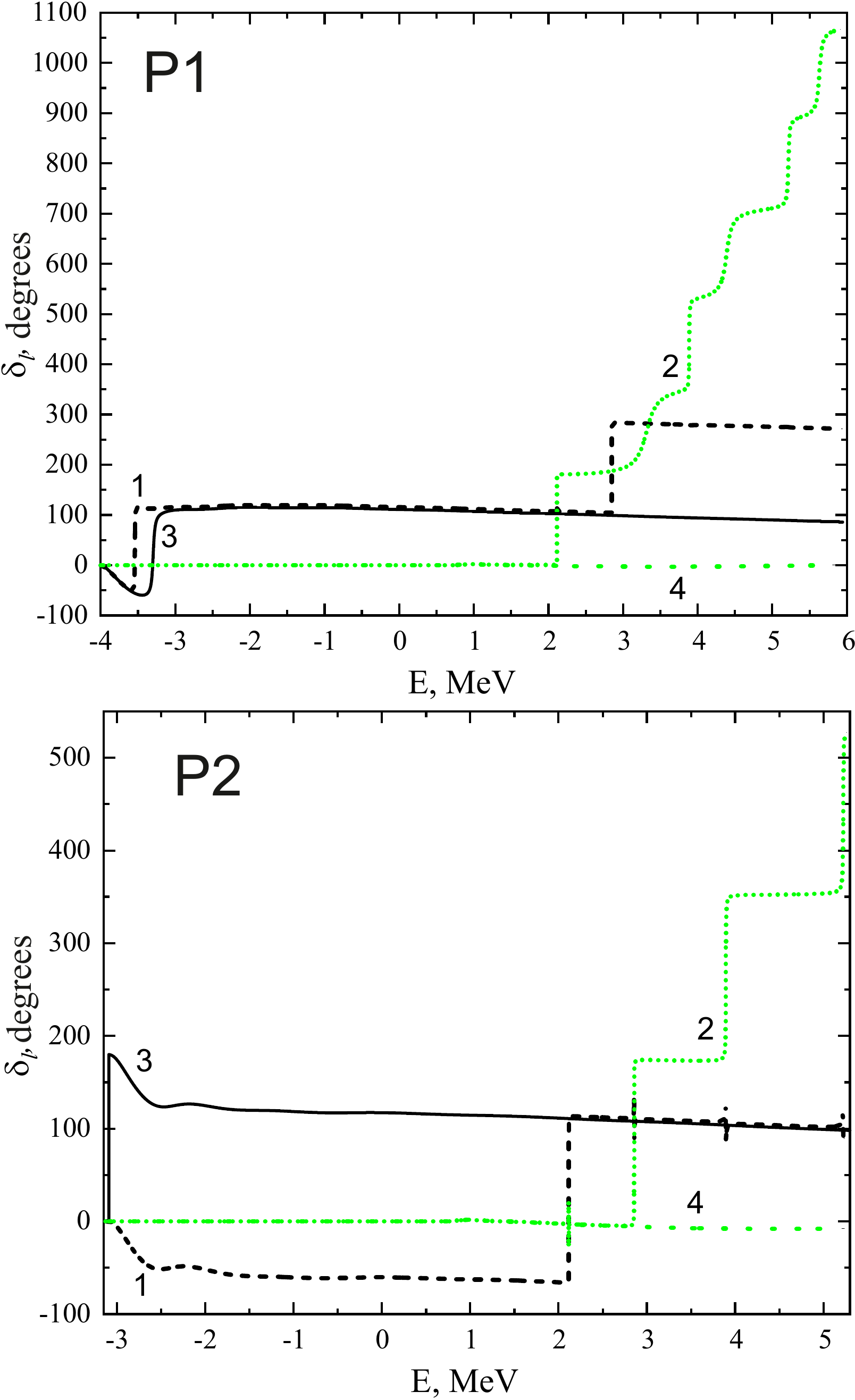}
\caption{Phase shifts of elastic many-channel scattering for $J^\pi(L)=3/2^+(2)$ states of $^9_\Lambda$Be versus energy in
the $_{\Lambda}^{5}$He$(0^+_1)$+$^{4}$He (1, 3) and $^{8}$Be$(0^+_1)$+$\Lambda$  (2, 4) channels. The data are obtained with $P1$ (upper panel) and $P2$ (lower panel) set of input parameters for soft (1, 2) and rigid (3, 4)
two-cluster subsystems. Orbital momenta of the relative motion of clusters $l_1$ and $l_2$ equals 2.}%
\label{Fig:delta_32+}%
\end{center}
\end{figure}

Figure \ref{Fig:delta_32+} illustrates the behaviour of the phase shifts of elastic
scattering for $J^\pi(L)=3/2^+(2)$ states of $^9_\Lambda$Be hypernucleus for two sets of the input parameters.
We can observe from Fig. \ref{Fig:delta_32+} a large impact of cluster polarization
on the phase shifts of elastic scattering. It is interesting to note that the phase
shift of $_{\Lambda}^{5}$He$(0^+_1)$+$^{4}$He scattering calculated with $P1$ input parameters manifests resonance behaviour
only for the lowest $3/2^+_1$(2) state and the $3/2^+_4$(2). All other resonances in $P1$ calculations become
apparent only in the behaviour of $^{8}$Be$(0^+_1)$+$\Lambda$ scattering phase.

\subsubsection{Spectrum of  $^9_\Lambda$Be for $P2$ set of the input parameters}
\label{subsubsec:P2}


The spectrum of positive and negative parity  states for $1/2\leq J\leq 7/2$  ($0\leq L\leq 4 $) of
the $^9_\Lambda$Be nucleus calculated  within the AMGOB model  for $P2$ set of the input parameters is presented in Tables \ref{Tab:Spectrum9Be+Gob_new}, \ref{Tab:Spectrum9Be-Gob_new}. Only 41 keV above the $^5_\Lambda$He+$\alpha$ decay threshold appears a narrow $1/2^+ (0)$ resonance with the width of 9 keV. This state was absent in $P1$ calculations. 

\begin{figure}[hptb]
\begin{center}
\includegraphics[width=\columnwidth]{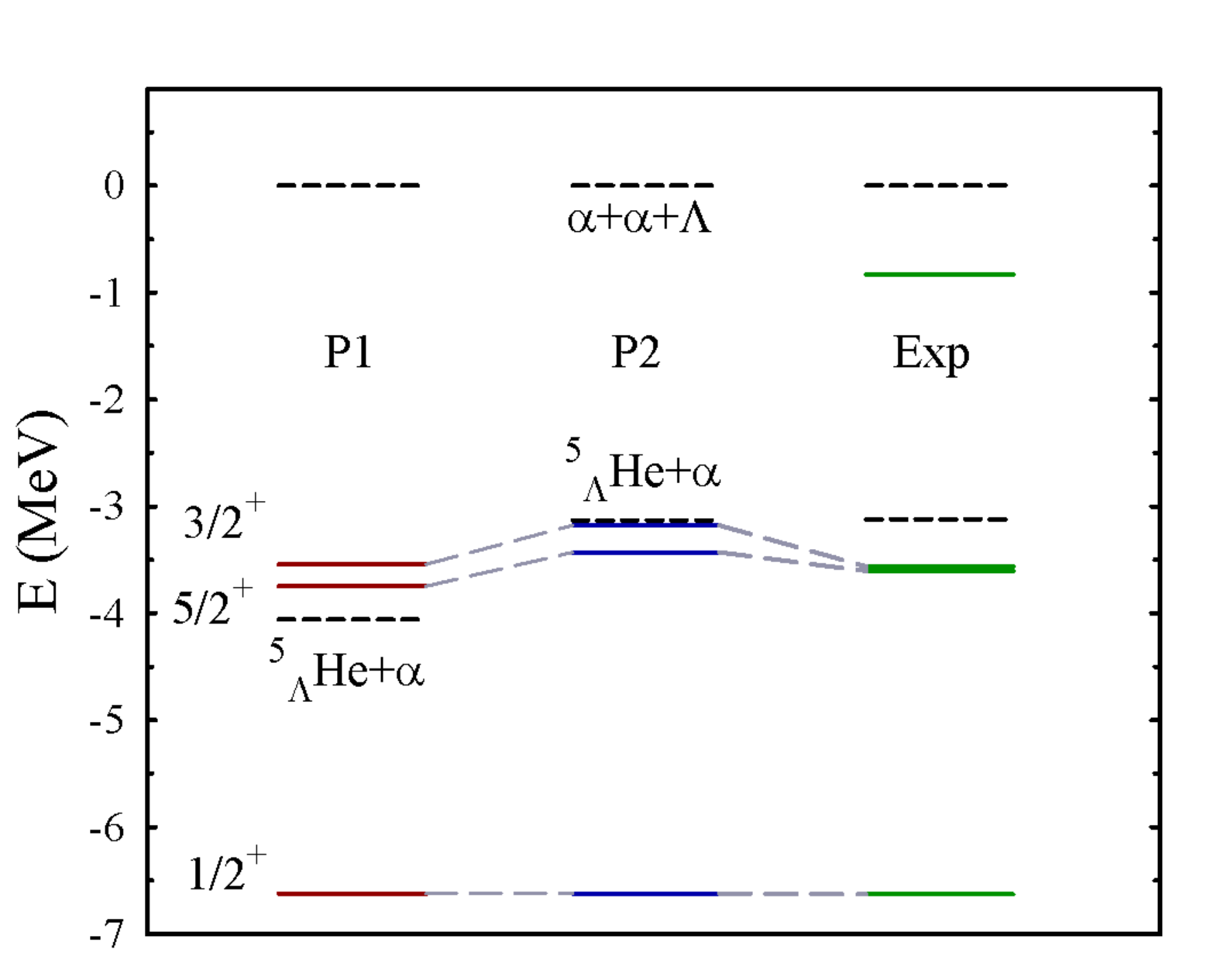}%
\caption{Energy of bound and resonance states in $_{\Lambda}^{9}$Be\ obtained
with two sets of input parameters $P1$ and $P2$ and compared to the experimental data (Exp) from Refs. \cite{2005NuPhA.754...58T, 2002PhRvL..88h2501A} }%
\label{Fig:Spectr9LBeP1P2}%
\end{center}
\end{figure}

In Fig. \ref{Fig:Spectr9LBeP1P2} we compare spectra of bound and resonance
states of the hypernucleus $_{\Lambda}^{9}$Be which lie below the
$\alpha+\alpha+\Lambda$ threshold and are obtained with $P1$  and
$P2$  sets of input parameters. These spectra are calculated with cluster polarization.  One sees that the $P2$ set of input
parameters shifts noticeable the energy of the $_{\Lambda}^{5}$He+$\alpha$ threshold and slightly shifts up energies of 3/2$^{+}$ (2) and 5/2$^{+}$ (2)
states. And thus these states are transformed into the bound states lying close to the
$_{\Lambda}^{5}$He+$\alpha$\ threshold. Neglecting cluster polarization turns bound 3/2$^{+}$ (2) state into a virtual state, while 5/2$^{+}$ (2) remains to be bound with respect to the $_{\Lambda}^{5}$He+$\alpha$ threshold.

Comparing the data presented in Tables \ref{Tab:Spectrum9Be+Gob_new}, \ref{Tab:Spectrum9Be-Gob_new} with those listed in Tables \ref{Tab:Spectrum9Be+Gob}, \ref{Tab:Spectrum9Be-Gob} we observe that $P2$ set of input parameters generates half as many resonances as $P1$ set. Besides, the majority of the resonances presented in Tables \ref{Tab:Spectrum9Be+Gob_new}, \ref{Tab:Spectrum9Be-Gob_new} decay mainly via $^5_\Lambda$He+$\alpha$ channel, while $P1$ set of parameters produced the resonances decaying into $^8$Be+$\Lambda$ channel. Furthermore, almost all resonances (except $7/2^+(4)$ case), which disappeared in $P2$ calculations, have larger widths and are characterized with partial width $\Gamma_2\gg\Gamma_1.$ The $7/2^+(4)$ wide resonance appeared to be almost insensitive both to the different sets of input parameters and to the accounting of cluster polarization. All narrow $7/2^+(4)$ resonances did not appear in $P2$ calculations.

\begin{table}[htbph] \centering%
\caption{Spectrum of positive parity states of $^9_\Lambda$Be nucleus obtained
within the AMGOB model with $P2$ input parameter set taking into account cluster polarization (soft) or neglecting
it (rigid). $\Gamma$ is a total width of the resonance state,
$\Gamma_{1,2}$ are partial decay widths of the resonance via $^5_\Lambda$He$+\alpha$
and $^8$Be$+\Lambda$ channels, correspondingly. Energy is given in MeV, the
total and partial widths are in keV.}%
\begin{tabular}{ccccccccc}
\hline
 & \multicolumn{4}{c}{rigid} & \multicolumn{4}{c}{soft}  \\ \hline
$J^{\pi} (L)$ & $E$ & $\Gamma $ & $\Gamma_1$ & $\Gamma_2$ & $E$ & $\Gamma $  & $\Gamma_1$ & $\Gamma_2$\\ \hline
$1/2^{+} (0)$& -6.33 &  &   &  &-6.62 &  &  & \\ \hline
         & -3.01 & 95 &  95     &      & -3.089 & 9   & 9  & \\ 
                &  &  &       &      & 1.82 & 13.9  & 6.7& 7.2 \\ 
                &  &  &       &      & 2.391 & 49.2 & 21.2 & 28 \\ 
                &  &  &       &      & 3.309 & 91.2 & 47.5 & 43.7 \\ 
                &  &  &       &      & 4.482 & 84.7 & 74.7 & 10\\                 
                &  &  &       &      & 4.5   & 85.5 & 83 & 2.5\\ \hline
$3/2^{+} (2)$       & -3.09 & 0.001  & 0.001      &      & -3.179 &   &  & \\ 
                &  &  &       &      & 2.116  & 0.36 & 0.22 & 0.14 \\ 
                &  &  &       &      & 2.856 &  2.84 & 10.2& 18.2\\ 
                &  &  &       &      & 3.895 & 10.3 & 3.9 & 6.4\\ 
                &  &  &       &      & 5.22 & 18.2 & 8 & 10.2\\ \hline
$5/2^{+} (2)$       & -3.34 &  &       &      & -3.43 &   &  & \\ 
                &  &  &       &      & 2.116  & 0.35 & 0.21 & 0.14\\ 
                &  &  &       &      & 2.856 &  2.70 & 0.93 & 1.77\\ 
                &  &  &       &      & 3.894  & 9.87 & 3.67 & 6.2\\ 
                &  &  &       &      & 5.22 & 17.7 & 7.6 & 10.1\\ \hline                
$7/2^{+} (4)$       & 4.67& 2295.1 & 2295.09& 0.02   & 4.658 & 2283  & 2282.98 & 0.02\\ \hline
\end{tabular}%
\label{Tab:Spectrum9Be+Gob_new}
\end{table}
\begin{table}[htbph] \centering%
\caption{Spectrum of negative parity states of $^9_\Lambda$Be nucleus obtained
within the AMGOB model with $P2$ input parameter set taking into account cluster polarization (soft) or neglecting
it (rigid). $\Gamma$ is a total width of the resonance state,
$\Gamma_{1,2}$ are partial decay widths of the resonance via $^5_\Lambda$He$+\alpha$
and $^8$Be$+\Lambda$ channels, correspondingly. Energy is given in MeV, the
total and partial widths are in keV.}%
\begin{tabular}{ccccccccc}
\hline
 & \multicolumn{4}{c}{rigid} & \multicolumn{4}{c}{soft}  \\ \hline
$J^{\pi}(L)$ & $E$ & $\Gamma $ & $\Gamma_1$ & $\Gamma_2$ & $E$ & $\Gamma $  & $\Gamma_1$ & $\Gamma_2$\\ \hline
$1/2^{-} (1)$       &  &  &       &      & 1.934 & 7.37  & 6.32 & 1.05\\ 
                &  &  &       &      & 2.574 & 36.29 & 31.29& 5\\ 
                &  &  &       &      & 3.554 & 75.1  & 68.8 & 6.3\\ 
                &  &  &       &      & 4.837 & 85.47 & 84.84 & 0.63 \\ \hline
$3/2^{-} (1)$       &  &  &       &      & 1.934 & 7.46  & 6.83 & 0.63\\ 
                &  &  &       &      & 2.574 & 35.78 & 31.44& 4.34\\ 
                &  &  &       &      & 3.554 & 73.68 & 68.44& 5.24\\ 
                &  &  &       &      & 4.837 & 85.59 & 85.31 & 0.28 \\ \hline
$5/2^{-} (3)$       &  &  &       &      & 2.345 & 0.06  & 0.059 & 0.001\\ 
                &  &  &       &      & 3.228 & 0.13  & 0.07 & 0.06\\ 
                &  &  &       &      & 4.40 & 0.74   & 0.36 & 0.38 \\ \hline
$7/2^{-} (3)$       &  &  &       &      & 2.345 & 0.083 & 0.082 & 0.001\\ 
                &  &  &       &      & 3.228 & 0.126 & 0.072 & 0.054\\ 
                &  &  &       &      & 4.40 & 0.72   & 0.37  & 0.35  \\ \hline
\end{tabular}%
\label{Tab:Spectrum9Be-Gob_new}
\end{table}
Energies of the $_{\Lambda}^{5}$He bound and pseudo-bound states obtained for $P2$ set of input parameters tabulated in Table \ref{Tab:2CSystems_new}. We can observe from Table \ref{Tab:2CSystems_new} that the energy of the bound state corresponds to its experimental value, while all the pseudo-bound states have much higher energies than in $P1$ calculations. Probably, this means that polarization of the $^5_\Lambda$He subsystem hardly affects the spectrum of the $^9_\Lambda$Be hypernucleus,  resulting in a smaller number of resonances in the $^9_\Lambda$Be for the $P2$ calculations.
\begin{table}[tbp] \centering
\caption{Energy and mass root-mean-square radius of the  $_{\Lambda}^{5}$He=$\alpha+\Lambda$ bound and
pseudo-bound states obtained for $P2$ set of input parameters. The energy is measured from the two-cluster threshold
indicated in the first column.}\vspace{2 mm}
\begin{tabular}
[c]{cccccc}\hline
2C-system & Quantity & $\sigma=1$ & $\sigma=2$ & $\sigma=3$ & $\sigma
=4$\\\hline
$_{\Lambda}^{5}$He=$\alpha+\Lambda$ & $E$, MeV & -3.13 &57.99 & 300.07 &
1013.10\\
& $r_{m}$, fm & 1.53 & 1.52 & 1.33 & 1.28\\ \hline
\end{tabular}
\label{Tab:2CSystems_new}%
\end{table}%

\subsubsection{Comparison with other methods and experiment}

In Table \ref{Tab:DiffMethods} we display
the spectrum of  $_{\Lambda}^{9}$Be obtained in Refs. \cite{2019FBS...60...30L}
and \cite{2000FBS....28..103O}, and compare it with our results.
\begin{table}[htbph] \centering
\caption{Spectrum of bound and resonance states in $^{9}_{\Lambda}$Be obtained in different models. Energy is displayed in MeV, width is given in keV.}%
\begin{tabular}
[c]{cccccccccccc}\hline
\multicolumn{3}{c}{\cite{2019FBS...60...30L}} &
\multicolumn{3}{c}{Present paper, P1} & \multicolumn{3}{c}{Present paper, P2} &
\multicolumn{3}{c}{\cite{2000FBS....28..103O}}\\\hline
$L^{\pi}$ & $E$ & $\Gamma$ & $J^{\pi}$ & $E$ & $\Gamma$ & $J^{\pi}$ & $E$ & $\Gamma$ &
$J^{\pi}$ & $E$ & $\Gamma$\\\hline
0$^{+}$ & -6.65 & - & ${1\over2}^{+}$ & -6.62 & - &  ${1\over2}^{+}$ & -6.62 & - & ${1\over2}^{+}$ & -7.13 & -
\vspace{1mm}\\
2$^{+}$ & -3.82 & - & ${5\over2}^{+}$ & -3.75 & 0.75 & ${5\over2}^{+}$ & -3.43 & - & ${5\over2}^{+}$ &
-3.89 & - \vspace{1mm}\\
 &  &  & ${3\over2}^{+}$ & -3.54 & 5.5 & ${3\over2}^{+}$ & -3.18 & - & ${3\over2}^{+}$ &
-3.89 & - \vspace{1mm}\\
1$^{-}$ & 0.1 & 2500 & ${1\over2}^{-}$ & 0.724 & 4228 & ${1\over2}^{+}$ & -3.09 &  9 & ${3\over2}^{-}$ &
-2.94 & 120\vspace{1mm}\\
 &  &  & ${3\over2}{-}$ & 1.92 & 4.0  & ${3\over2}^{-}$ & 1.93 & 7.46 & ${3\over2}^{-}$ & -2.21 &
2160\\
 &  &  & ${3\over2}^{-}$ & 3.37 & 258.6  & ${3\over2}^{-}$ & 3.55 & 73.68 & ${3\over2}^{-}$ &
3.86 & 5.6\vspace{1mm}\\
4$^{+}$ & 3.2 & 780& ${7\over2}^{+}$ & 2.61 & 0.01 & ${7\over2}^{+}$ & 4.66 & 2283 &  &  & \vspace{1mm}\\
3$^{-}$ & 8.0 & 6100& ${7\over2}^{-}$ & 2.35 & 0.15 & ${7\over2}^{-}$ & 2.35 & 0.08 &  & \vspace{1mm}\\ \hline
\end{tabular}
\label{Tab:DiffMethods}%
\end{table}%


As this nucleus has not more than two bound states and a large number of
resonance states, we selected those investigations which determined both
energies and widths of the excited states.

In Ref.
\cite{2019FBS...60...30L} a three-cluster model with an approximate treatment
of the Pauli principle was employed.  The $\alpha\Lambda$ interaction \ was
determined in a folding approximation with the same $\Lambda N$ potential as
in the present paper, however another value of the parameter $k_{F}$
($k_{F}$ = 0.963 fm$^{-1}$) was selected. The complex scaling method was
involved in Ref. \cite{2019FBS...60...30L} to locate the energy and width of
resonance states.

As the spin-orbit interaction is disregarded in Ref.
\cite{2019FBS...60...30L}, the authors labelled the states of the $^9_\Lambda$Be
with the total orbital momentum $L$. The energy of the ground state, obtained
in Ref. \cite{2019FBS...60...30L} is close to our results.  There is also a certain
likeness of our results for both input parameter sets and the results of Ref. \cite{2019FBS...60...30L} for the energy
of the $2^{+}_1$ state. However,
in our approach this state is split by spin-orbit interaction on two states -
5/2$_{1}^{+}$(2) and 3/2$_{1}^{+}$(2). In $P1$ calculations these states are resonances, since they are located above the lowest $^{5}_\Lambda$He$\left(0_{1}^{+}\right)+\Lambda$
threshold, while with $P2$ input parameters these states are bound. An important point is that our model reproduce the correct order
of the 5/2$_{1}^{+}$(2) and 3/2$_{1}^{+}$(2) states and only slightly overestimates their splitting for both sets of the input parameters.

 The $L^\pi=1^-$ resonance state presented in Ref. \cite{2019FBS...60...30L} corresponds to the lowest $1/2_1^-(1)$ 
resonance obtained in our model with $P1$ set of the input parameters. Both states are located close to the $2\alpha+\Lambda$ and have a large width.

There are some differences in the position and in the widths of other resonance states. The energy of the $4^+$ resonance listed in Ref.\cite{2019FBS...60...30L} is about 0.6 MeV higher the energy of the lowest $7/2^+(4)$ resonance obtained with $P1$ input parameter set of our model and a significantly larger width. At the same time, $7/2^+(4)$ resonance produced by $P2$ input parameter set is characterized by higher energy and larger width than the $4^+$  resonance from Ref. \cite{2019FBS...60...30L}. Finally, the lowest $7/2^-(3)$ resonance state produced by both sets of input parameters in our model have essentially lower energy and smaller width than the $3^-$ resonance presented in Ref. \cite{2019FBS...60...30L}.


The Faddeev equation methodology were applied in Ref.
\cite{2000FBS....28..103O} to study bound and resonance states in $_{\Lambda
}^{9}$Be within a macroscopic three-body model. Several effective
$\alpha\Lambda$ interactions \ and a separable $\alpha\alpha$ interaction
were involved to calculate the spectrum.
The selected $\alpha\Lambda$ \ and $\alpha\alpha$ interactions lead to
the overbound ground state of $^9_\Lambda$Be in Ref. \cite{2000FBS....28..103O}.
As the spin-orbit interaction is neglected in \cite{2000FBS....28..103O},
the 3/2$^{+}$ and 5/2$^{+}$ have the same energy. They are bound states, since
they are  located under the $_{\Lambda}^{5}$He+$\alpha$
threshold. Three $3/2^{-}$ resonance states were found in Ref.
\cite{2000FBS....28..103O}, two of them are located bellow the three-cluster
threshold and one above.  $P2$ set of the input parameters of our model produces the lowest $1/2_1^+(1)$ resonance with the energy close to the energy of the lowest $3/2^-$ resonance from Ref.
\cite{2000FBS....28..103O}.
The $3/2^{-}_3$ state from \cite{2000FBS....28..103O}
lies near our 3/2$^{-}$ resonance states with energy 3.37 MeV and 3.55 MeV for $P1$ and $P2$ input parameter set, correspondingly. The latter states have, however, larger widths  than the width of the 3/2$_{3}^{-}$ resonance
state obtained in Ref. \cite{2000FBS....28..103O}.

In Table \ref{Tab:Spectrum9BeExp} we collected experimental data on the spectrum of the $^9_\Lambda$Be hypernucleus. The references are found in Ref. \cite{2019FBS...60...30L}.
\begin{table}[htbph] \centering%
\caption{Spectrum of $_{\Lambda}^{9}$Be identified in different experiments compared with our results. Energy is given in MeV}%
\begin{tabular}{ccccccccc}
\hline
Source & \cite{1983PhRvL..51.2085M,
1976PhLB...62..481B} & \cite{2006PrPNP..57..564H,
 1998NuPhA.639...93A} & \multicolumn{2}{c}{\cite{2005NuPhA.754...58T%
, 2002PhRvL..88h2501A}} & \multicolumn{2}{c}{Theory, P1} &\multicolumn{2}{c}{Theory, P2}\\ \hline
$J^{\pi }$ & $E$  & $E$  &$J^{\pi }$& $E$  &$J^{\pi }$& $E$ &$J^{\pi }$& $E$
\\ \hline
1/2$^{+}$ & -6.63 & -5.90 & 1/2$^{+}$ & -6.62 & 1/2$^{+}_1$ & -6.62 & 1/2$^{+}_1$ & -6.63 \\ 
 &  &  & 5/2$^{+}_1$ & -3.606 & 5/2$^{+}_1$ & -3.75 & 5/2$^{+}$ & -3.43   \\ 
3/2$^{+}$& -3.55 &  -2.97 &  3/2$^{+}$ & -3.563 &  3/2$^{+}_1$ & -3.54 &  3/2$^{+}_1$ & -3.18  \\ 
&  & -0.10  &  &-0.83 & 1/2$_{1}^{-}$ & 0.724 & 1/2$_{2}^{+}$ & -3.09\\ 
&  &  &  &2.89 & 5/2$^{+}_3$ & 2.849 &  5/2$^{+}_3$ & 2.856\\ 
&  &  &  &  &  3/2$^{+}_3$ & 2.85& 3/2$^{+}_3$ & 2.856\\ 
&  &  3.62 &  &  &  3/2$^{-}_4$ & 3.37 & 3/2$^{-}_3$ & 3.55\\\hline
\end{tabular}%
\label{Tab:Spectrum9BeExp}%
\end{table}
As we can see from Table \ref{Tab:Spectrum9BeExp}, there is some consistency
in experimental data only for the energies of the ground $1/2^+$ state and
excited $3/2^+$, $5/2^+$ states.
The energies of these three levels obtained within our model are in a good
agreement with the experimental data from the first and last column. Energy of the lowest $3/2_1^-$ state obtained with $P1$ input parameter set is closer to the experimental value than that produced in $P2$ calculations. However, in the latter case $3/2_1^-$ and $5/2_1^-$ are located below the $^5_\Lambda$He$+\alpha$ decay threshold, in accordance with the experimental data, while in $P1$ calculations these states lie above the two-cluster threshold. Energies of the $3/2_3^+$ and $5/2_3^+$ resonance states are close to the experimentally observed level with $E=2.89$ MeV from Ref. \cite{2005NuPhA.754...58T, 2002PhRvL..88h2501A}. Finally, the resonance state with energy $E=3.62$ MeV reported in \cite{2006PrPNP..57..564H, 1998NuPhA.639...93A} is located near our $3/2_4^-$ and $3/2_3^-$ resonances generated by $P1$ and $P2$ input parameter sets, correspondingly.

\subsection{Nature of the resonance states}

As we can see above, our model generates a large number of resonance states in
$_{\Lambda}^{9}$Be. What is the nature of such resonance states? What factors
are responsible for the appearance of these resonance states?   All the results of this Subsection have been obtained with $P1$ set of the input parameters, since this variant of calculations produces more resonance states.

There are at
least three possible reasons for the formation of resonance states and thus three
types of resonances emerged in our many-channel model of $_{\Lambda}^{9}$Be.
First, the so-called shape resonance \ states can be created by a centrifugal or the Coulomb
barriers, or by a combination of both barriers.

The centrifugal barrier is present in both $_{\Lambda}^{5}$He+$\alpha$
and $^{8}$Be+$\Lambda$ channels, if the total orbital momentum and parity
allow rotational states with nonzero orbital momentum of relative motion of
the interacting clusters. The Coulomb barrier is present in the $_{\Lambda}^{5}$He+$\alpha$
channel only. Effects of the Coulomb forces on  parameters of the resonance
states are demonstrated in Table \ref{Tab:CoulombRS32P} with $J^\pi=3/2^{+}$ states.
This example shows us that the Coulomb interaction is not responsible for creating
such a rich variety of resonance states. As we have seen, it changes parameters of
a resonance state and transforms a weakly bound state into a resonance state.

Second, some part of the obtained resonance states may be considered
as the Pauli resonance states. Appearance of a large number of
resonance states in cluster systems has been discovered many years ago
(see, for example, Refs.
\cite{1969PhRv..179..971T, 1973PhRvC...8.1649T, 1974PhLB...49..308C,
1982PhRvC..26.2410S,1982NuPhA.380...87K, 1983NuPhA.394..387W,
1985NuPhA.437..367W, 1988PhRvC..38.2013K}).
In Refs. \cite{1982NuPhA.377...84F, 1975PThPh..54..747K} it has been
shown with a a simple two-cluster model systems how the Pauli resonances appear
in the resonating group method calculations when
different oscillator lengths are used for the interacting clusters. It is worthwhile
noticing that with distinct values of oscillator lengths for different clusters one achieves more adequate description of a
compound system. The Pauli resonance states may appear also in the case when the
same oscillator lengths are used but more advanced internal cluster functions are adopted.
Using different oscillator lengths implies invoking monopole excitations of each cluster. Such excitations in light nuclei have the energy around 20 MeV. And, consequently, the Pauli resonances appear in a high-energy range. Meanwhile, employment of more advanced wave functions of interacting clusters by considering them as a binary structure may result in a set of low-energy internal states which generate the Pauli resonances at a low energy. That is the case for the present investigation.

The Pauli resonances is caused by the almost Pauli forbidden states, which
can be present in the wave function instead of the Pauli forbidden states
both in the case of different oscillator lengths and advanced intrinsic cluster wave function.
Recall that the forbidden states are the eigenfunctions of the norm
kernel with zero eigenvalues, while the almost Pauli forbidden states
correspond to nonzero but very small eigenvalues.
In Ref.
\cite{1992NuPhA.548...39K}\ it was claimed that the Pauli resonances
disappear if the almost forbidden states are removed from the wave function.
It was demonstrated for the $^{16}$O+$\alpha$ scattering that by omitting the
eigenstates with eigenvalues less than 1.0$\times$10$^{-2}$ \ one removes the
Pauli resonances. We apply this algorithm to calculate the 3/2$^{+}$(2) phase
shifts and parameters of the resonance states. In this case we have got only one
almost forbidden state with the eigenvalue 6.4$\times$10$^{-8}$. Other
eigenvalues are spread in the region between 0.2 and 1.8. Elimination of the
almost forbidden state did not result in changing phase shifts and
parameters of the 3/2$^{+}$(2) resonances. The same results were obtained
for the 1/2$^{+}$(0) and 1/2$^{-}$(1) states. This let us conclude that the obtained
resonances are not the Pauli resonances. We suggest that the Pauli resonance
states arise at a higher energy.

Third, Feshbach resonances appear in many-channel systems, provided that the
channels have different threshold energies (see original papers of Feshbach \cite{1958AnPhy...5..357F,1962AnPhy..19..287F}, a review in Ref. \cite{2010RvMP...82.1225C} and Ref. \cite{2019PTEP.2019l3D02M}, where the Feshbach resonance state was discovered in $^{11}$Li considered as a three-body system $^9$Li$+n+n$). The Feshbach resonance
originates from a bound states in one of the channels with a higher energy of the decay threshold.
Having coupled with other channel(s) with
lower threshold energy, the bound state may then decay through the open
channel(s). If the coupling between channels is weak, the bound
state turns into a resonance state. If the coupling is strong, a bound state can
be dissolved by continuum.
We have  the necessary conditions for
creation of the Feshbach resonance states, since the binary channels involved
in the calculations have different threshold energies (see Table
\ref{Tab:2CSystems}). In the region 1 MeV$<$E$<$6 MeV, where we discovered the majority of the resonance states, there are four closed channels which can generate bound states.

However, to prove that our resonance states are the Feshbach resonances, we need to
compare spectrum of the resonance states with the spectrum of bound states in closed binary channels treated separately. To obtain spectrum of the bound states, we
selected from a huge matrix of the Hamiltonian those blocks which describe
each channel and omitted the blocks which couple them. Thus problem
of the $N_{ch}$ coupled channels are reduced to the problem of $N_{ch}$
independent channels and each channel is treated separately. As we pointed
above, there is a weak coupling between different channels of $_{\Lambda}^{9}$Be. Thus
we expect a correspondence between energies of the bound and resonance states.

\begin{figure}[ptb]
\begin{center}
\includegraphics[width=\columnwidth]{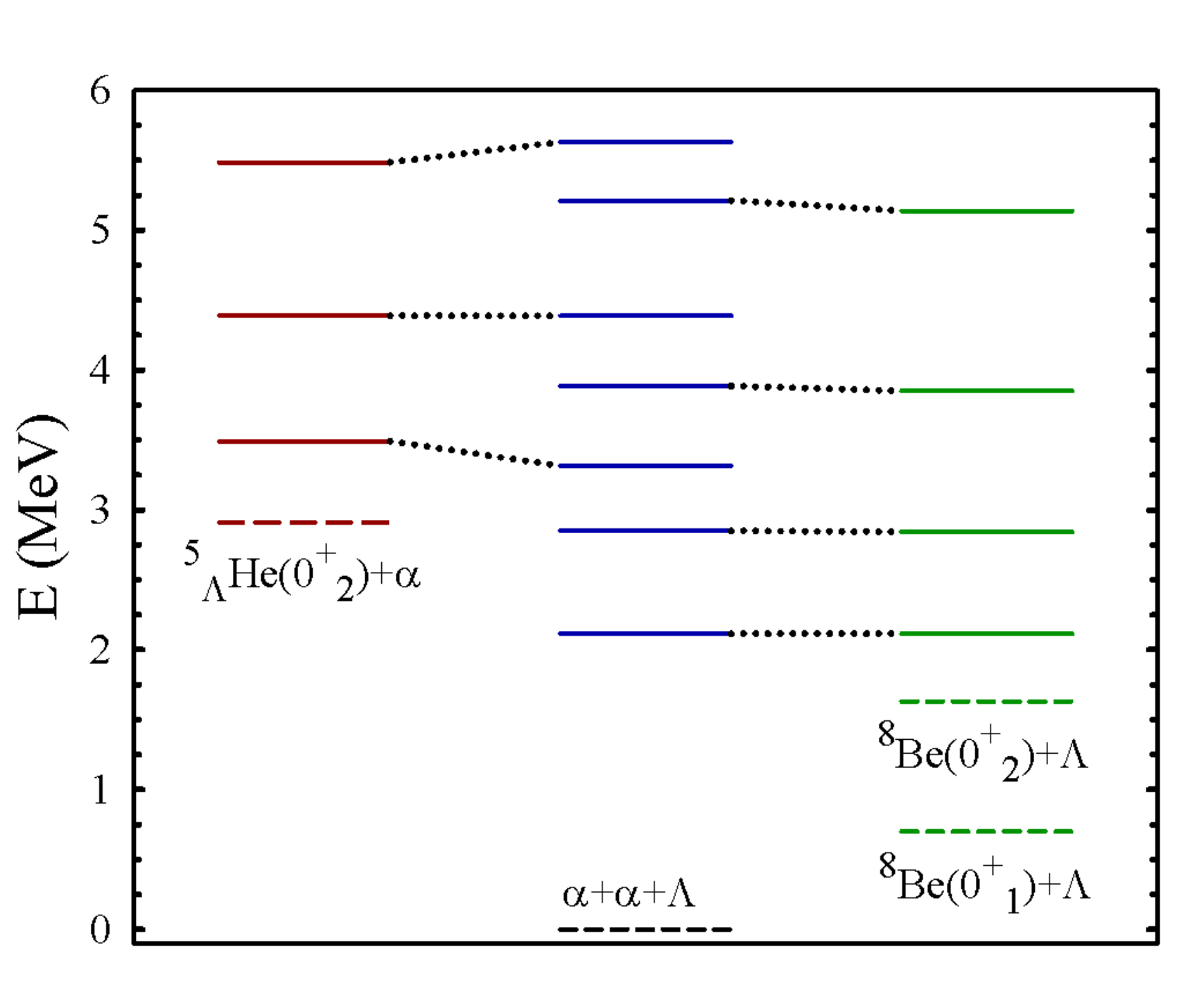}
\caption{Spectrum of the 3/2$^{+}$(2) resonance states in $^9_\Lambda$Be obtained in the coupled-channel approximation compared with the bound states  created by closed binary channels $_{\Lambda}^{5}$He$\left(0_{3}^{+}\right)
$+$\alpha$ (the left spectrum) and  $^{8}$Be$\left(  0_{3}^{+}\right)  $+$\Lambda$ (the right spectrum) treated separately. $P1$ set of the input parameters was used.}%
\label{Fig:Feshbach32P}%
\end{center}
\end{figure}

Comparison is made in Fig. \ref{Fig:Feshbach32P}\ where we display the energy
of the 3/2$^{+}$(2) resonance states created by the eight coupled channels (the middle spectrum)
with the energy of bound states in the channel $_{\Lambda}^{5}$He$\left(0_{3}^{+}\right)
$+$\alpha$ (the left spectrum) and channel $^{8}$Be$\left(  0_{3}^{+}\right)  $+$\Lambda$ (the right spectrum).
 We display only those bound states in all closed
channels which lie in the energy range of our interest.
Spectrum of bound states in each channel is obtained with 200 oscillator
functions. By the dashed lines we indicated two- and three-body thresholds.

One can see in Fig. \ref{Fig:Feshbach32P} that the energies of the resonance
states obtained in the coupled-channel approximation are very close to the
energies of the bound states calculated within a single-channel approximation.
It is necessary to recall that the threshold energies of the channels $%
_{\Lambda }^{5}$He$\left( 0_{3}^{+}\right) +\alpha $ and $^{8}$Be$\left(
0_{3}^{+}\right) +\Lambda $ equal 21.12 and 9.11 MeV, correspondingly, as it
was indicated in\ Table \ref{Tab:2CSystems}. And thus all energy levels, which are
obtained in the single-channel approximations and lay below these threshold
energies, are bound states.
The $_{\Lambda}^{5}$He$\left(  0_{3}^{+}\right)  $+$\alpha$ closed channel
generates three resonance states, while the $^{8}$Be$\left(0_{3}^{+}\right)$+$\Lambda$
closed channel  creates four resonance states. Thus, the
numerous resonance states emerged from our calculations are mainly the
Feshbach resonance states. They complement the shape resonance states created
by the centrifugal and Coulomb barriers which appear at the low-energy region.
From Fig. \ref{Fig:Feshbach32P} one may conclude that the coupling of open channels with the  $_{\Lambda}^{5}$He$\left(  0_{3}^{+}\right)  $+$\alpha$ is stronger than with the $^{8}$Be$\left(0_{3}^{+}\right)$+$\Lambda$, since the resonance states are shifted noticeably with respect to the corresponding bound states in the first channel but not with respect to the states of the second channel.

 Summarizing, we can conclude that the location of narrow resonances of
the $^9_\Lambda$Be hypernucleus in the energy region above $^8$Be$(0_1^+)+\Lambda$
threshold depends on the pseudo-bound states of the $^8$Be and $^5_\Lambda$He
two-cluster subsystems, which have been taken into account. We can assume that
the experimentally observed levels of the two-body subsystems of the considered
hypernucleus should be primarily accounted for. For the $^9_\Lambda$Be hypernucleus
these are $0^+$ ground state and $2^+$, $4^+$ resonance states of the $^8$Be nucleus,
as well as the $0^+$ ground state of the $^5_\Lambda$He hypernucleus. In the present paper, we properly accounted for the ground states of both two-body subsystems in
the $^9_\Lambda$Be hypernucleus. In addition, we allowed for both subsystems to be
polarized without changing their angular momenta. There are strong grounds to believe that
the location of the resonances of the $^9_\Lambda$Be located below $^8$Be$(0_1^+)+\Lambda$
threshold remains unaffected by considering $2^+$ and $4^+$ pseudo-bound states of the $^8$Be
subsystem instead of already taken into account $0^+_2$, $0^+_3$ and $0^+_4$ states. However,
it can change the energies of the $^9_\Lambda$Be resonance states lying at the energy range
$E>2$ MeV above the three-cluster threshold.

\section{Conclusions}
\label{sec:concl}

We have applied a microscopic three-cluster model to studying the structure of the $_{\Lambda}^{9}$Be hypernucleus. The model treats correctly the Pauli principle and accounts for polarization of two-cluster subsystems of the hypernucleus when the third cluster is close by reducing a three-cluster problem to a coupled many-channel two-cluster problem.
A Gaussian basis was used for the description of the two-cluster subsystems, while a harmonic oscillator basis was invoked for the relative motion of the two-cluster subsystem and the remaining cluster.

The hypernucleus $_{\Lambda}^{9}$Be was considered as a three-cluster system
comprising of two alpha particles and a $\Lambda$-hyperon. Within the present
model the three-cluster configuration was projected on two binary
configurations $_{\Lambda}^{5}$He+$\alpha$ and $^{8}$Be+$\Lambda$, provided that $_{\Lambda}^{5}$He and $^{8}$Be were described as two-cluster systems. A finite number of Gaussian functions were used to describe $\alpha-\Lambda$ and $\alpha-\alpha$ binary subsystems. This resulted in a discretization of the continuous spectrum of $_{\Lambda}^{5}$He and $^{8}$Be. The set of pseudo-bound states in $_{\Lambda}^{5}$He and $^{8}$Be allowed us to take into account polarizability of these systems.

The spectrum of bound and resonance states in the  $_{\Lambda}^{9}$Be hypernucleus was studied in detail. Calculations have been performed with two set of input parameters. Both sets were selected to reproduce the ground state energy of $_{\Lambda}^{9}$Be and to minimize the energy of the $^8$Be two-cluster subsystem of this hypernucleus. At the same time, $P2$ input parameter set was determined to reproduce the binding energy of $_{\Lambda}^{5}$He with respect to its $\alpha+\Lambda$ decay threshold, while $P1$ input parameter set minimizes binding energy of $_{\Lambda}^{5}$He making it slightly overbound.

It was shown that the cluster polarizability plays a significant role in formation of bound and resonance states of this hypernucleus. Moreover,  polarization of the two-cluster subsystems on interaction with the third cluster is responsible for creation of  a large number of resonance states, a large part of them are very narrow  with the total width less than 100 keV. We have shown that the majority of the narrow resonances in the energy range 1 MeV$<$E$<$6 MeV are the Feshbach resonances generated due to a weak coupling of different binary channels in $^9_\Lambda$Be.  $P2$ set of the input parameters generate the less number of resonance states than $P1$ set due to higher energies of pseudo-bound states in $_{\Lambda}^{5}$He.

There is a fairly good agreement between our results and available experimental
data. However, in our calculations with the first set of input parameters $P1$, the first 3/2$^{+}$(2) excited state of the $^9_\Lambda$Be is a resonance state, since it is located above the lowest $_{\Lambda}^{5}$He+$\alpha$ decay threshold of the $^9_\Lambda$Be. This can be ascribed to the selected nucleon-nucleon and nucleon-hyperon potentials which slightly overbind the $^5_\Lambda$He subsystem.
 At the same time, $P2$ set of the input parameters make the first 3/2$^{+}$(2) and  5/2$^{+}$(2) states to be bound in accordance with the experiment. The order of lowest $5/2^+$(2) and $3/2^+$(2) levels corresponds to the experimental data for both sets of the input parameters. There is a very good correspondence of the predicted by our model energies of the $3/2_3^+$ and $5/2_3^+$ resonance states to the experimentally observed level with $E=2.89$ MeV from Ref. \cite{2005NuPhA.754...58T, 2002PhRvL..88h2501A}.

\section{Acknowledgment}

This work was supported in part by the Program of Fundamental Research of
the Physics and Astronomy Department of the National Academy of Sciences of
Ukraine (Project No. 0117U000239).

\end{document}